\documentclass[fleqn,10pt]{wlscirep}

\usepackage[utf8]{inputenc}
\usepackage[T1]{fontenc}
\usepackage{amsmath,amssymb,amsbsy,amsxtra,graphicx}
\usepackage{array,multirow}
\usepackage{booktabs}
\usepackage{url}
\usepackage{cite}
\usepackage{bm}
\usepackage{caption}
\usepackage{subfig}
\usepackage{epstopdf}
\usepackage{color}
\usepackage{balance}
\usepackage{graphicx}

\usepackage{float}
\usepackage[flushleft]{threeparttable}
\usepackage{dblfloatfix}
\usepackage{comment}
\usepackage[T1]{tipa}
\usepackage{booktabs,array}
\usepackage{adjustbox}

\title{Automatic vocal tract landmark localization from midsagittal MRI data}

\author[1,+]{Mohammad Eslami}
\author[1]{Christiane Neuschaefer-Rube}
\author[1,*]{Antoine Serrurier}
\affil[1]{Clinic for Phoniatrics, Pedaudiology \& Communication Disorders, University Hospital and Medical Faculty, RWTH Aachen University, Germany.}
\affil[+]{meslami@ukaachen.de}
\affil[*]{aserrurier@ukaachen.de}

\begin{abstract}
The various speech sounds of a language are obtained by varying the shape and position of the articulators surrounding the vocal tract. Analyzing their variations is crucial for understanding speech production, diagnosing speech disorders and planning therapy.
Identifying key anatomical landmarks of these structures on medical images is a pre-requisite for any quantitative analysis and the rising amount of data generated in the field calls for an automatic solution.
The challenge lies in the high inter- and intra-speaker variability, the mutual interaction between the articulators and the moderate quality of the images.
This study addresses this issue for the first time and tackles it by means by means of Deep Learning.
It proposes a dedicated network architecture named \textit{Flat-net} and its performance are evaluated and compared with eleven state-of-the-art methods from the literature.
The dataset contains midsagittal anatomical Magnetic Resonance Images for 9 speakers sustaining 62 articulations with 21 annotated anatomical landmarks per image.
Results show that the \textit{Flat-net} approach outperforms the former methods, leading to an overall Root Mean Square Error of 3.6 pixels/0.36 cm obtained in a leave-one-out procedure over the speakers. The implementation codes are also shared publicly on GitHub.

\end{abstract}
\begin{document}

\flushbottom
\maketitle

\thispagestyle{empty}

% #########################
% #########################
% #########################
% #########################
% #########################
% #########################
% #########################
% #########################

\section{Introduction}
\label{sec-introduction}

In speech, the sounds of a language are produced by varying the shape and position of the organs surrounding the vocal tract.
% This region is characterized by a high inter- and intra-speaker articulatory variability, both in the space and time domains.
This region is characterized by a high inter- and intra-speaker variability, both in the space and time domains.
Analyzing and modeling the shape of the vocal tract articulators is therefore crucial for speech production research \cite{harshman1977,beautemps2001,s_serrurier2019b} and for diagnosis and therapy of related disorders, including speech disorders \cite{yamasaki2017vocal, guzman2017computerized}, velopharyngeal insufficiency \cite{VPI} and swallowing dysfunctions \cite{disorders}.
% The vocal tract extends from the glottis to the lips and comprises various structures such as the larynx, including the epiglottis, the velum, the tongue, and the upper and lower lips. 
The vocal tract area extends from the glottis to the lips and is surrounded by various structures such as the larynx, the epiglottis, the velum, the tongue, and the upper and lower lips.
Despite the high variability, the organisation of these structures is similar among all speakers.
Measuring and analyzing their variability implies therefore to know the deformation of these similar structures across speakers and articulations. A pre-requisite to achieve this is to match similar pertinent anatomical features on these structures across speakers and articulations.
% A pre-requisite for analyzing their variability is to match pertinent anatomical features on these structures across speakers and articulations.
% The key component to achieve this is to identify on each articulation the same pertinent anatomical landmarks characterizing these structures or demarcating them.
The key component for this purpose is to identify on each articulation the same pertinent anatomical landmarks characterizing these structures or demarcating them.
It constitutes the general framework of the study.

Articulatory speech production studies often rely on midsagittal images of the vocal tract area and Magnetic Resonance Imaging (MRI) constitutes in this approach an essential modality \cite{MriSeg, narayanan2004, story2005b}. Identifying landmarks of the vocal tract area on these images has always been done manually or as a byproduct of manual segmentation \cite{echternach2016, s_serrurier2019b}. If such a manual approach fulfilled the needs until nowadays, the exploding number of data in this field due to the rise of real-time MRI \cite{review-vt-apps} and the recent progress in data science call undoubtedly for an automatic approach. This study aims at solving this issue and is to our knowledge the first study to address it.
% This study focuses therefore on the localization of anatomical landmarks of the vocal tract on midsagittal MRI.
% Nowadays, landmarks are labelled manually on MRI data \cite{echternach2016, s_serrurier2019b}. However, the increasing number of data \cite{review-vt-apps} and the recent progress in data science call for an automatic approach.
% To our knowledge, this is the first study to consider this issue and to propose a solution to solve it automatically.

This objective takes place in a larger framework in biomedical engineering and computer vision where localizing anatomical landmarks on biomedical images, sometimes referred to as detecting keypoints, has already been considered in other contexts.
% Localizing anatomical landmarks on biomedical images, sometimes referred to as detecting keypoints, has already been considered for other structures and applications.
% As an illustration, this has for instance been considered for the aortic valve on Computer Tomography scans \cite{valve}, for cephalometric landmarks on lateral cephalograms \cite{Cephalometric, geometric}. Detecting finger joints in X-ray and MRI and Vertebrae locations in volumetric CT scans of the spine  are two more examples \cite{heatmap_2019}.
As an illustration, this has for instance been considered for the aortic valve on Computer Tomography scans \cite{valve}, for cephalometric landmarks on lateral cephalograms \cite{Cephalometric, geometric}, for finger joints on X-ray and MRI or for spine landmarks on volumetric Computer Tomography scans \cite{heatmap_2016,heatmap_2019}.
It has also been considered as a key element of more global registration processes, such as between fundus photographs and MRI for the eye \cite{eye} and between series of biological microscopic images \cite{microscope}.
Regions of interests have also been identified by means of landmark localization, such as for the brain for Alzheimer's disease diagnosis \cite{AD1} or to detect changes in facial temperature \cite{face-temperature}.
% It has also been used to identify regions of interest, such as for the brain for Alzheimer's disease diagnosis \cite{AD1} or to detect changes in facial temperature \cite{face-temperature}.
One can finally also mention the use of landmark identification for research on non-human animals \cite{cat}.
% WE CAN ADD MORE HERE.
This short review emphasizes the importance of anatomical landmark localization from images for a large variety of applications and this study lengthens this non-exhaustive list with speech production.

% While landmark localization is an important task in medical image analysis, these applications take place in a more general framework in computer vision where localizing automatically landmarks on images is a long-standing problem.
% Two fields of application in this context appear more particularly active and connected to our problem.
Within the existing literature, two fields of application appear more particularly active and connected to our problem.
% The first is the localization of landmarks for the face, for which a comprehensive review is provided by Wu \textit{et al.}\cite{facial-survey}.
The first is the localization of landmarks for the face, rich of an abundant research from landmark identification on two-dimensional photographs \cite{HyperFace,DAN} to landmark identification on three-dimensional Ultrasounds for fetuses for instance \cite{ultrasound_1,ultrasound_2}. A recent and  comprehensive review is provided by Wu \textit{et al.}\cite{facial-survey}.
% Facial landmark detection represents a challenging issue due to the high variability of the shapes, poses, occlusions, and lighting conditions.
It represents a challenging issue due to the high variability of the shapes, poses, occlusions, and lighting conditions.
% A current challenge is the detection of the 68 landmarks specified in the \textit{300W} dataset \cite{300w}.
Similarly, localizing the position of the joints of the body on images to estimate the human pose is also a long-standing problem \cite{pose-survey}. It is also a challenging issue in computer vision due to the high variability of the postures, body shapes, actions, clothes and scenes.
% A current challenge is the detection of the 16 body joint positions specified in the \textit{MPII Human Pose} dataset \cite{andriluka20142d}.
% These problems constitute the closest problems found in the literature related to our issue and actively studied so that a wide range of approaches are proposed.
% The solution proposed in this this study are therefore inspired by this literature. 

% One of the state-of-the-art methods to localize facial landmarks relies on regressions implemented in a boosting approach, the \textit{dlib} method \cite{dlib}. The majority of the recent techniques rely however on deep learning (DL) neural networks, known to be powerful to automatically identify and combine key features from input data to solve the considered problem. Notable current methods for facial landmark estimation include the so-called \textit{deep alignment network} \cite{DAN} and the \textit{HyperFace} method \cite{HyperFace}. For human pose estimation, a recent method is the \textit{multi-context attention model} \cite{MCAM}, an extension of the \textit{stacked hourglass networks} \cite{Stacked-hourglass}. The architecture of the networks used in these methods are based on Convolutional Neural Networks (CNNs), particularly adapted to the processing of images. 

The goal and contribution of this study is to propose a fully-automated end-to-end image analysis methods for localizing key anatomical landmarks in the vocal tract area from midsagittal MRI data.
% The goal and contribution of this study is to investigate and propose fully-automated end-to-end image analysis methods for localizing key anatomical landmarks of the vocal tract from midsagittal MRI data.
As emphasized earlier, the need for such a method in the field is required and has never been attempted so far.
It is aimed to be used in the future for new speakers for which no prior data are available.
Considering the existing methods of the literature and the recent rise of data science to solve such problems, Deep Learning (DL) appears as the inescapable approach\cite{dl}. Indeed, DL approaches in image processing tasks appear to outperform most of traditional techniques while bringing more robustness to noise\cite{dl_image}.
In image processing, Deep Neural Networks usually take the particular form of Convolutional Neural Networks (CNNs) \cite{dl_cnn}. This study intends therefore to provide a CNN architecture able to solve our specific problem. This method and the associated network will be referred to as \textit{Flat-net} in this paper.
As detailed in the section \ref{sec-method}, the design of this method is inspired from existing solutions listed earlier.
% For this reason, a method based on deep learning (DL) architecture is proposed in this study named \textit{Flat-net} which brings more accuracy and robustness among speakers.
It takes notably advantage of the concept of \textit{heat-maps} \cite{heatmap-origin}, which are images where the maximum intensities correspond to the landmark coordinates. It transforms interestingly coordinates in images, leading to input and output data of same nature, and appears powerful to deal with landmarks in image processing\cite{heatmap_2016, heatmap-origin}.

In addition, in an attempt to compare our solution with existing approaches found in the literature for landmark or joint localization in other contexts, eleven state-of-the-art methods have been adapted to our problem and implemented as our competitors: (1) four methods from the facial landmark localization literature, (2) one method from the human pose estimation literature and (3) six generic methods from the literature on landmark localization in medical images and based on \textit{heat-maps}.
% In addition, eleven available methods in literature are considered as competitors too. 
% Four methods from facial landmark localization and one method from human pose estimation, all designed for the localization of landmarks/joints on images, will be adapted to our problem and evaluated. 
% Moreover, six solutions based on heat-map regression are also considered which are DL networks and basically Convolutional Neural Networks (CNNs). 
% a dedicated method specifically designed for the localization of vocal tract landmarks on midsagittal MRI will be proposed.

% These methods aim at describing each single landmark as a full image with a maximal intensity on the landmark localization, referred to as a \textit{heat-map}. 
% Each landmark can be described as a heat-map places in channels of a single image, leading to the concept of \textit{heat-maps in channels}. 
% Therefore localization of landmarks consists therefore in generating \textit{heat-maps in channels} from input MRI image. It is worth to mention that the proposed \textit{Flat-net} is also based on \textit{heat-map} regression.
% Seven DL networks will be tested for the generation of such heat-maps: (1) six from literature known to be efficient to generate heat-maps as in our problem, and (2) one dedicated network developed on purpose in this study (\textit{Flat-net}).

One of the challenges of DL approaches lies in the very large amount of data necessary for training. Standardised datasets are publicly available for classical problems such as facial landmark detection or body pose estimation \cite{300w,andriluka20142d}.
No benchmark exists for the problem described in this study and datasets in the field of articulatory speech analyses are usually rather limited and characterized by high shape and noise variability \cite{s_serrurier2019b}.
% The dataset considered for this study is in line with this observation and therefore much more limited and heterogeneous than the datasets mentioned above for facial landmark localization and human pose estimation, increasing all the more the challenge.
The dataset considered for this study is in line with this observation and therefore much more limited and heterogeneous than the datasets mentioned above, increasing all the more the challenge.
However, preliminary analyses \cite{ESSV} taking advantage of DL algorithms for image segmentation carried out on this exact dataset proved the feasibility of such an approach.
% The code for the  methods considered and evaluated in this study is shared publicly for research and validation purposes (seven by us and five by others).
The code for the  methods considered in this study is publicly available, either by ourselves or by other research groups, for research and validation purposes.

The rest of the paper is organized as follows: in section \ref{sec-method}, the dataset, the anatomical landmarks, the methods, and the evaluation schemes are presented; the section \ref{sec-results} reports the experimental results; finally, the methodology and the results are discussed in the section \ref{sec-discussion}, together with the perspectives.

% #########################
% #########################
% #########################
% #########################
% #########################
% #########################
% #########################
% #########################

\section{Methods}
\label{sec-method}

\subsection{Data}

The study considers static midsagittal MRI recorded between 2002 and 2011 from 9 French speakers (5 males, 4 females), referred to as \textit{subjects} in this study, sustaining 62 different articulatory positions, also referred to as \textit{classes} in the context of machine learning, designed to be representative of the French phonemic repertoire \cite{staticMRIData, s_serrurier2019b}. 
This study does not include any human experiments more than the use of non-invasive MRI data collection mentioned above. 
All the methods and data acquisition were carried out in accordance with the guidelines and regulations of the local ethic committee called CPP, 'Comité de Protection des Personnes' \cite{CPP} (English translation: Committee for the Protection of People) and the recording protocols were approved by this ethic committee.
All subjects were older than 18 and an informed consent was obtained from all of them. 
The images have been recorded either on a 1.5 or on a 3 Tesla MRI scanner and have a field of view of $256\times256$ mm$^2$ and a resolution of 1 mm per pixel.
Note that two speakers have been discarded in comparison to \cite{staticMRIData, s_serrurier2019b} due to the significantly lower quality of the images, the different fields of view and the different sizes of the images.

\begin{figure}[h]
\centering
  \includegraphics[width=0.35\linewidth, trim={6cm 10cm 6cm 10cm},clip]{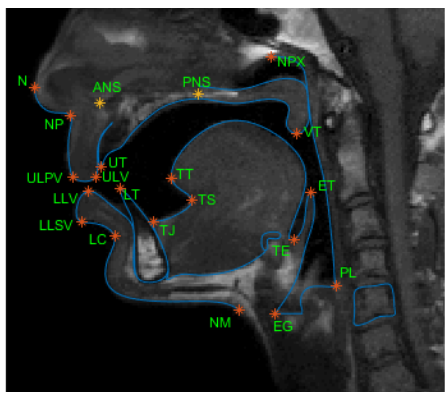}
  \caption{(Color online) MRI superimposed with the 21 anatomy landmarks of the study.}
  \label{fig:landmarks}
\end{figure}

21 anatomical landmarks relevant to the study of the speech articulations have been identified. They represent either characteristic landmarks of the speech articulators, such as the tip of the tongue, or the junction between two articulators. They are listed in Table \ref{tab:landmarks} and illustrated for one articulation of one subject in Fig. \ref{fig:landmarks}. They have been manually identified on all images of the dataset by an expert.
Note that the upper and lower teeth landmarks (UT and LT) denote dental structures and, as such, are not distinguishable from the air on MRI data.
They have been determined for each subject by contrast with soft tissues on an articulation acquired on purpose and reported on the other images using their relative position with the hard palate. Please refer to \cite{s_serrurier2019b} for further information regarding this procedure as well as the data collection and processing.

\begin{table*}[h]
\centering
\caption{List of the landmarks of interest for vocal tract area MRI image analysis.}

\begin{adjustbox}{width=\textwidth}

\begin{tabular}{|c|c|c|}
\hline
\textbf{Abbreviation} & \textbf{Name}                 & \textbf{Description}                                                                    \\ \hline
ANS                   & Anterior Nasal Spine          & Anterior nasal spine                                                                    \\ \hline
EG                    & Epiglottis-Glottis            & Junction between the epiglottis and the glottis                                         \\ \hline
ET                    & Epiglottis Tip                & Tip of the epiglottis                                                                   \\ \hline
LC                    & Lip-Chin                      & Labiomental groove                                                                      \\ \hline
LLSV                  & Lower Lip Skin Vermillion     & Vermillon border of the lower lip                                                       \\ \hline
LLV                   & Lower Lip Vermillion          & Junction between the wet and dry vermillion of the lower lip                            \\ \hline
LT                    & Lower Teeth                   & Upper point of the lower incisors                                                       \\ \hline
N                     & Nose                          & Most anterior point of the tip of the nose                                              \\ \hline
NM                    & Neck-Mandible                 & Junction between the horizontal submandibular line and vertical neck line \\ \hline
NP                    & Nose-Philtrum                 & Junction between the philtrum and the external nose                                     \\ \hline
NPX                   & Nasopharynx                   & Upper point of the nasopharynx                                                          \\ \hline
PL                    & Pharynx-Larynx                & Junction between the  pharyngeal wall and the posterior supraglottic region             \\ \hline
PNS                   & Posterior Nasal Spine         & Posterior nasal spine                                                                   \\ \hline
TE                    & Tongue-Epiglottis             & Junction between the tongue and the epiglottis                                          \\ \hline
TJ                    & Tongue-Jaw                    & Junction between the tongue and the jaw                                                 \\ \hline
TS                    & Tongue Sub                    & Most posterior point of the sublingual cavity                                           \\ \hline
TT                    & Tongue Tip                    & Tip of the tongue                                                                       \\ \hline
ULPV                  & Upper Lip Philtrum Vermillion & Vermillon border of the upper lip                                                       \\ \hline
ULV                   & Upper Lip Vermillion          & Junction between the wet and dry vermillion of the upper lip                            \\ \hline
UT                    & Upper Teeth                   & Lower point of the upper incisors                                                       \\ \hline
VT                    & Velum Tip                     & Tip of the velum                                                                        \\ \hline
\end{tabular}

\end{adjustbox}
\label{tab:landmarks}
\end{table*}

\subsection{Challenges}
\label{subsec-challenges}

\begin{figure*}[ht]
\captionsetup{justification=centering}
  \includegraphics[width=\linewidth]{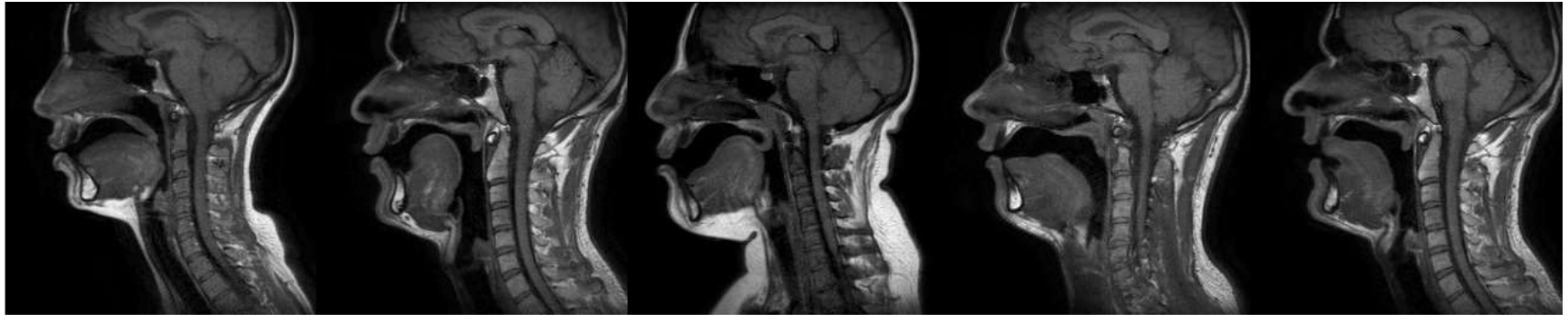}
  % \caption{Showing some of the challenges of the landmarks in vocal tract environment including touching articulations, various shapes and deformations, sizes, blurriness and invisible tooth.}
  \caption{Five images of the database illustrating the variability.}
  \label{fig:challenges}
\end{figure*}

Localizing landmarks of the vocal tract area on midsagittal MRI images presents particular characteristics. Fig. \ref{fig:challenges} shows a few articulations from different subjects illustrating the diversity of the dataset.
The main challenges are summarized in the following list.

\begin{itemize}
    \item The shapes and positions of the vocal tract articulators are characterized by a high variability, due to the variety of the speech task and to the different morphologies and articulatory strategies of the speakers to perform a same task.
    \item Some articulators such as the tongue, velum, lips and epiglottis present a high variability in the vocal tract area, leading to very different locations in the dataset associated to the landmarks VT, TT, TS, LLSV, ULPV, ET and TE.
    \item Some tissues may touch each others for certain articulations, leading to hardly distinguishable landmarks at these locations, such as TE, TS and ET.
    \item The larynx area appears very difficult to capture on midsagittal MRI data, leading to a confusing area on the images and hardly identifiable landmarks, such as PL and EG.
    \item Some articulators may occasionally show very different shapes than in the large majority of the articulations, such as the velum rolled up against the tongue for a few articulations \cite{staticMRIData}, leading to unusual location of the landmark VT.
    \item Two important landmarks for speech production analyses are the teeth landmarks UT and LT, which are not directly visible on the images, as mentioned earlier.
    \item The images are recorded at different times with different scanners, leading to variable quality and noise levels. Similarly, the quality and noise level may not be homogeneous within a single image.
    \item Despite the high variability, the size of the available dataset, \textit{i.e.} 9 subjects with 62 articulations, hence 558 articulations, is rather limited in comparison to those usually used for landmark localization in the literature, \textit{e.g.} over 10K images for mentioned references \cite{HyperFace, RCN, andriluka20142d}.
\end{itemize}

\subsection{Methods}
\label{sec-methods}

% Twelve different methods are compared.
% Eleven methods are state-of-the-art methods taken from the literature and adapted to the current problem (four for detection of 68 facial landmarks, one for detection of 16 body joints and six for detection of anatomical landmarks by heat-map regression). One method named \textit{Flat-net} is the dedicated method introduced in this paper. 
In order to compare the method proposed in this paper with the current state-of-the-art in the field, eleven methods taken from the literature have been adapted to our problem and implemented: (1) five methods from the literature related to facial landmark detection and body pose estimation and (2) six general methods for landmark localization and based on the concept of \textit{heat-maps}. Our method, referred to as \textit{Flat-net}, is inspired by these methods and presented at the end.

\subsubsection{Methods from facial landmark detection and body pose estimation}
% The methods regarding to facial landmarks and body joints are \textit{dlib}, \textit{HyperFace}, \textit{Deep Alignment Network}, \textit{Shape Fitting by Deep-Regression} and \textit{Multi-Context Attention Model} as follow.
% The methods regarding to facial landmarks and body joints are \textit{dlib}, \textit{HyperFace}, \textit{Deep Alignment Network}, \textit{Shape Fitting by Deep-Regression} and \textit{Multi-Context Attention Model} as follow.
The five methods considered in this section are the \textit{dlib}, \textit{HyperFace}, \textit{Deep Alignment Network}, \textit{Shape Fitting by Deep-Regression} and \textit{Multi-Context Attention Model} methods and are presented as follows.

\subsubsection*{\textbf{\textit{dlib}} }
The algorithm available as part of the \textit{dlib} library is an implementation of the ensemble of regression trees presented in 2014 by Kazemi and Sullivan \cite{dlib}.
This technique takes advantage of simple features with fast computing capacities, \textit{e.g.} the pixels' intensity differences, to directly estimate the landmark locations.
These locations are subsequently refined with an iterative process made of a cascade of regressors and using gradient boosting.
Note that the \textit{dlib} method is the only method considered in this study not based on DL. 

\subsubsection*{\textbf{\textit{HyperFace}}}
The \textit{HyperFace} method makes use of an end-to-end DL network for simultaneous face detection, landmark localization, pose estimation and gender recognition \cite{HyperFace}.
It exploits the intermediate layers of a deep CNN, such as the \textit{ResNet-101} \cite{ResNet}, by connecting together the intermediate feature maps to further predict the various desired outputs.
The last layers of the \textit{HyperFace} network are fully connected layers.

\subsubsection*{\textbf{\textit{Deep Alignment Network}}}

The \textit{Deep  Alignment  Network (DAN)} is a method based on a DL to localize facial landmarks \cite{DAN}.
It consists of multiple stages of CNNs, where each stage improves the locations of the facial landmarks estimated by the previous stage.
A key element of the system is the use of \textit{heat-maps} within each stage.
In their approach, a \textit{heat-map} is defined as an image with highest intensity values at the exact locations of all the considered landmarks and decreasing intensities around as a function of the distance to the nearest landmark. The last two layers of each stage are fully connected layers.

\subsubsection*{\textbf{\textit{Shape Fitting by Deep-Regression}}}
% \textit{Shape Fitting Deep-Regression (SFD)} is proposed recently and combines deep convolutional neural networks with model-based fitting algorithms such as PCA of the landmark positions \cite{shapenet}. The PCA is included in the deep neural network using a novel layer type. This network predicts very fast and can detect landmarks at several hundreds of frames per second.
The \textit{Shape Fitting Deep-Regression (SFD)} method has been proposed recently. It combines CNNs with model-based fitting algorithms of the landmark positions such as obtained by Principal Component Analysis (PCA) \cite{shapenet}. The PCA is included in the network using a dedicated layer type. This network is characterized by a fast computing performance and can process until several hundreds of frames per second.

\subsubsection*{\textbf{\textit{Multi-Context Attention Model}}}

The \textit{Multi-Context Attention Model (MCAM)} method is an extended version of the DL \textit{stacked hourglass networks} \cite{Stacked-hourglass} designed for human pose estimation and body joint localization \cite{MCAM}.
It generates \textit{heat-maps} describing the body joint locations by using multiple resolutions, conditional random fields and an original layer type combining various convolutional layers together.
For training, the ground truth \textit{heat-maps} are generated by 2-D Gaussians centered on the joint locations.
The generated \textit{heat-maps} contain all joint locations together and are further split into partial \textit{heat-maps} for each body joint by means of an extra spatial classifier.
Since the network is designed to localize 16 body joints, two of these networks are necessary in practice in the current study to localize the 21 landmarks.

\subsubsection{Heat-map-based methods}
% \textcolor{blue}{Alternatively in a different fashion, the deep learning networks for landmark localization are exploited  to predict the location of landmarks via creating the heat-maps. One of the pioneer attempt for this reason is the work by Pfister et al. \cite{heatmap-origin} to regress heat-maps for landmarks simultaneously instead of absolute landmark coordinates for body pose estimation.}
Previous methods aim at detecting landmark coordinates from images. One limitation of these approaches come from the different nature of the input and output data. An alternative to overcome this is to consider the landmark coordinates as images via \textit{heat-maps} in output. One of the pioneering attempt to implement such approach can be attributed to Pfister \textit{et al.} for landmark coordinate detection for body pose estimation \cite{heatmap-origin}.

In this approach, a single landmark is described as a full image, the \textit{heat-map}, with a maximal intensity on the landmark location.
Several landmarks can be considered on the same \textit{heat-map} or can be described as several channels of an image (or tensor), leading to the \textit{heat-maps in channels}(one \textit{heat-map} per landmark).
% Several landmarks are described as several channels of an image (or tensor), the \textit{heat-maps in channels}.
% Localizing the landmarks on an image consists therefore in generating the associated \textit{heat-maps in channels}, instead of the vectors of the landmarks' coordinates or one heat-map containing all landmarks as for the previous studies.
Localizing the landmarks on an image consists therefore in generating the associated \textit{heat-maps}, instead of the vectors of the landmarks' coordinates as for the previous studies.
% The landmark locations are then straightforwardly derived from the \textit{heat-maps in channels} as the points with maximal intensity in each channel.
The landmark locations are then straightforwardly derived from the \textit{heat-maps} as the points with maximal intensity in each channel.
This kind of approach appears particularly appropriate for medical image processing \cite{heatmap_2016,heatmap_2019}. 
Note that this concept of \textit{heat-map} is related to the concept of \textit{heat-maps} mentioned in the \textit{DAN} and \textit{MCAM} methods. However, in these methods, the \textit{heat-maps} contain all landmark/joint locations together and are not directly provided as output of the networks.

\begin{figure}[h]
\centering
\captionsetup{justification=centering}
  \includegraphics[width=0.5\linewidth]{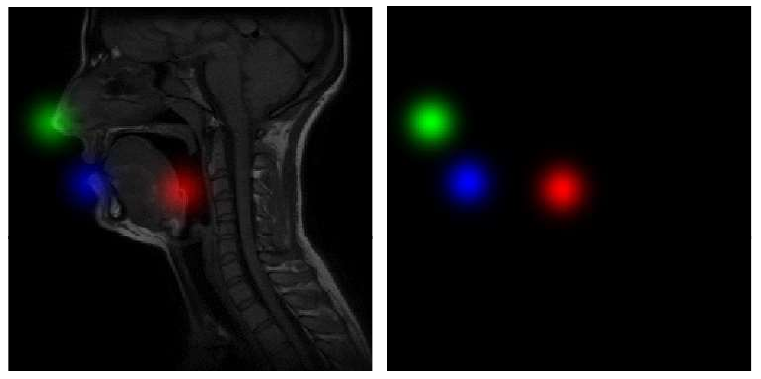}
  \caption{(Color online) Superposition of 3 \textit{heat-maps} for 3 different landmarks, represented as the 3 channels of a single RGB image (right), and superimposed on the corresponding input MRI image (left).}
  \label{fig:heatmaps}
\end{figure}

Practically in using \textit{heat-maps in channels}, for an input image of size $N \times N$, the network generates $L$ output \textit{heat-map} images (\textit{i.e.} tensor $H \in R^{N \times N \times L}$) where the $l^{th}$ \textit{heat-map} ($H_l \in R^{N \times N \times 1}$) has the maximum intensity value at the location of the estimated $l^{th}$ landmark. The normalized target \textit{heat-maps} are obtained via a Gaussian hat with a $\sigma$ (=10 in our experiment) pixels and a maximum of $1$ around each landmark location. Fig. \ref{fig:heatmaps} shows an example of the combination of 3 \textit{heat-maps} created for 3 landmarks and displayed as the 3 channels of a single RGB image. 
The predicted location of the $l^{th}$ landmark corresponds to the location of the maximum value in the $l^{th}$ predicted \textit{heat-map}:
\begin{equation}
    % (x_p,y_p)_l=argmax\{H_l\} ,  \ l=1,...,L 
   (x_p,y_p)_l=\operatorname*{argmax}_{x,y} {H_l(x,y)} ,  \ l=1,...,L 
\end{equation}

Six methods based on the generation of \textit{heat-maps} are considered in this section. The first five methods make use of the implementation proposed by Payer \textit{et al.} \cite{heatmap_2016} and for which the code is publicly available.
% Payer et. al. examine several deep learning architecture for medical landmark localization via heat-map generation including \textit{ConvOnly-net}, \textit{U-net}, and \textit{Downsampling-net} \cite{heatmap_2016}.
% They also propose a new architecture called \textit{SpatialConfiguration-net} and its extension with embedded \textit{U-net} \cite{heatmap_2016, heatmap_2019}.
Payer \textit{et al.} investigate several DL approaches for landmark localization on medical images based on the generation of \textit{heat-maps}. They implement existing networks such as \textit{Downsampling-net}, \textit{ConvOnly-net}, and \textit{U-net}.
They also propose a new architecture called \textit{SpatialConfiguration-net} and its extension with embedded \textit{U-net} \cite{heatmap_2016, heatmap_2019}.
% These five methods show promising results for localizing diverse landmarks on various biomedical image modalities such as finger landmarks on X-Ray and MRI, cranium landmarks on lateral cephalogram X-Ray and spine landmarks on Computer Tomography scans \cite{heatmap_2016,heatmap_2019}.
These five methods show promising results for localizing diverse landmarks on various biomedical image modalities as mentioned in the introduction \cite{heatmap_2016,heatmap_2019} (finger landmarks on X-Ray and MRI, cranium landmarks on lateral cephalogram X-Ray and spine landmarks on Computer Tomography scans).
These methods are briefly described below, please refer to the dedicated references \cite{heatmap_2016, heatmap_2019} for further details.
The last method presented in this section, the \textit{pix2pix’s generator network} covers our own implementation of the concept of \textit{heat-maps} into the existing \textit{pix2pix} network.

\subsubsection*{\textbf{\textit{Downsampling-net}}}
% \textit{Downsampling-net (DS-net)} exploits alternating convolution and pooling layers. Due to downsampling, this method can cover large image areas with small kernel sizes but because of low resolution target heat-maps, poor accuracy in localization has to be expected \cite{heatmap_2016}. In our experiment, this network with default hyper-parameters did not converge.
The \textit{Downsampling-net (DS-net)} methods alternates convolution and pooling layers in the network. The downsampling approach ensures covering large image areas with small kernel sizes but poor accuracy in localization has to be expected due to the associated low resolution of the resulting \textit{heat-maps} \cite{heatmap_2016}. This network with its default hyper-parameters did not converge in our experiment.

\subsubsection*{\textbf{\textit{ConvOnly-net}}}
% \textit{ConvOnly-net (conv-net) } is used to overcome the low target resolution via exploiting neither pooling layers, nor strided convolution layers \cite{heatmap_2016}. This network consists of 6 convolution layers.  
The \textit{ConvOnly-net (conv-net)} method aims at overcoming the low resolution of the generated \textit{heat-maps} mentioned above by discarding all pooling and strided convolution layers \cite{heatmap_2016}. This network results in six convolution layers.  

\subsubsection*{\textbf{\textit{U-net}}}
% \textit{U-net }\cite{u-net} is a well-known architecture for medical image analysis specially segmentation. \textit{U-net} includes convolutional layers with skip connections and pooling layers in an auto-encoder architecture which makes it to grasp large images even with small kernel size. The implementation by \cite{heatmap_2016} is slightly changed such as replacing the maximum pooling with average pooling. 
% which makes the width/heights of the output heatmaps same as input images. 
The \textit{U-net }\cite{u-net} is a well-known architecture for medical image processing, more specifically for segmentation. The \textit{U-net} includes convolutional layers with skip connections and pooling layers in an auto-encoder architecture, making efficient the processing of large images even with small kernel size.
% The implementation by \cite{heatmap_2016} is slightly changed such as replacing the maximum pooling with average pooling.
The implementation proposed by Payer \textit{et al.} \cite{heatmap_2016} and used in this study is a slightly modified version, where for instance the maximum pooling has been replaced by an average pooling. 

\subsubsection*{\textbf{\textit{SpatialConfiguration-net}}}
% \textit{SpatialConfiguration-net (SCN)} is proposed in the \cite{heatmap_2016} and includes three block architecture that combines local appearance of landmarks with the spatial configuration to all other landmarks. The first block of the network consists of three convolutional layers with small kernel sizes, that result in local appearance heat-maps. Second and third blocks are the spatial configuration block and combiner blocks which are designed to suppress locations from local appearance predictions that are infeasible due to the spatial configuration of landmarks \cite{heatmap_2016}.
The \textit{SpatialConfiguration-net (SCN)} method combines interestingly the local appearance of the landmarks with their spatial locations in reference to all other landmarks. It consists of three blocks. The first block is made of three convolutional layers with small kernel sizes, resulting in local appearance \textit{heat-maps}. The second and third blocks refine the outputs of the first block by considering the spatial configurations and by combining them with the initial outputs to discard the unrealistic results \cite{heatmap_2016}.

\subsubsection*{\textbf{\textit{SpatialConfiguration-net with embedded U-net}}}
% The \textit{SpatialConfiguration-net with embedded U-net (SCN(U-net))} method is an extension of the \textit{SCN} method which uses \textit{U-net} as an embedded network architecture for local appearance extraction block \cite{heatmap_2016, heatmap_2019}. 
The \textit{SpatialConfiguration-net with embedded U-net (SCN(U-net))} method is an extension of the \textit{SCN} method which embeds a \textit{U-net} network into the local appearance extraction block \cite{heatmap_2016, heatmap_2019}.

% In order to increase the accuracy of the vocal tract landmark localization, two more alternative methods are considered in this study.
% These two methods also exploit the concept of \textit{heat-maps in channels} and are implemented on two different DL networks, one designed on purpose for this study and the other taken from the literature: (1) the \textit{generator network of the pix2pix} method \cite{pix2pix}, extracted from the literature and adapted to the current needs and (2) the \textit{Flat-net} network, designed on purpose.

% In our approach, the output of the networks are directly the heat-maps, \textit{i.e.} images. 
% For this reason, the fully connected layers found in the \textit{HyperFace} and \textit{DAN} networks, transforming feature maps into landmark location vectors for output, can be omitted.

%\textit{b) \textbf{pix2pix's generator network:}}
\subsubsection*{\textbf{\textit{pix2pix's generator network}}}
This architecture exploits the generator component of the pix2pix network \cite{pix2pix}, referred to in the following as \textit{p2p-GN}, standing for \textit{pix2pix's generator network}.
It is based on a hourglass-shaped CNN with skip connections.
This network is specifically designed to analyze and generate images, hence particularly adapted in our case for the generation of \textit{heat-maps} from MRI. It has already proved to be very efficient for such tasks \cite{pix2pixHD}.
In the current study, this network is adapted to generate particular types of images, the \textit{heat-maps}.
The loss function as well as the hyper-parameters are those reported by \cite{pix2pix}.
Experimental analyses showed that this network produces better results when it does not generate more than 3 \textit{heat-maps} at the same time.
% For this reason, similarly to the \textit{Flat-net} approach, the prediction of the 21 heat-maps are split into 7 different networks. % Again, 
% Again, the additional number of weights resulting from the use of 7 networks in this approach is discussed in section \ref{sec-results}.
For this reason the prediction of the 21 \textit{heat-maps} are split into 7 different networks.
The additional number of weights resulting from the use of 7 networks in this approach is discussed in section \ref{sec-results}.

\subsubsection{Proposed network architecture}

This section details our own solution to solve the problem. It implements the concept of \textit{heat-maps-in-channels} described in the previous section and proposes a dedicated network for the generation of such \textit{heat-maps}.

%\textit{\textbf{flat-net network}:} 
\subsubsection*{\textbf{\textit{Flat-net}}}

\begin{figure*}[h]
\captionsetup{justification=centering}
  \includegraphics[width=\linewidth,trim={0cm 12.75cm 0cm 6.5cm},clip]{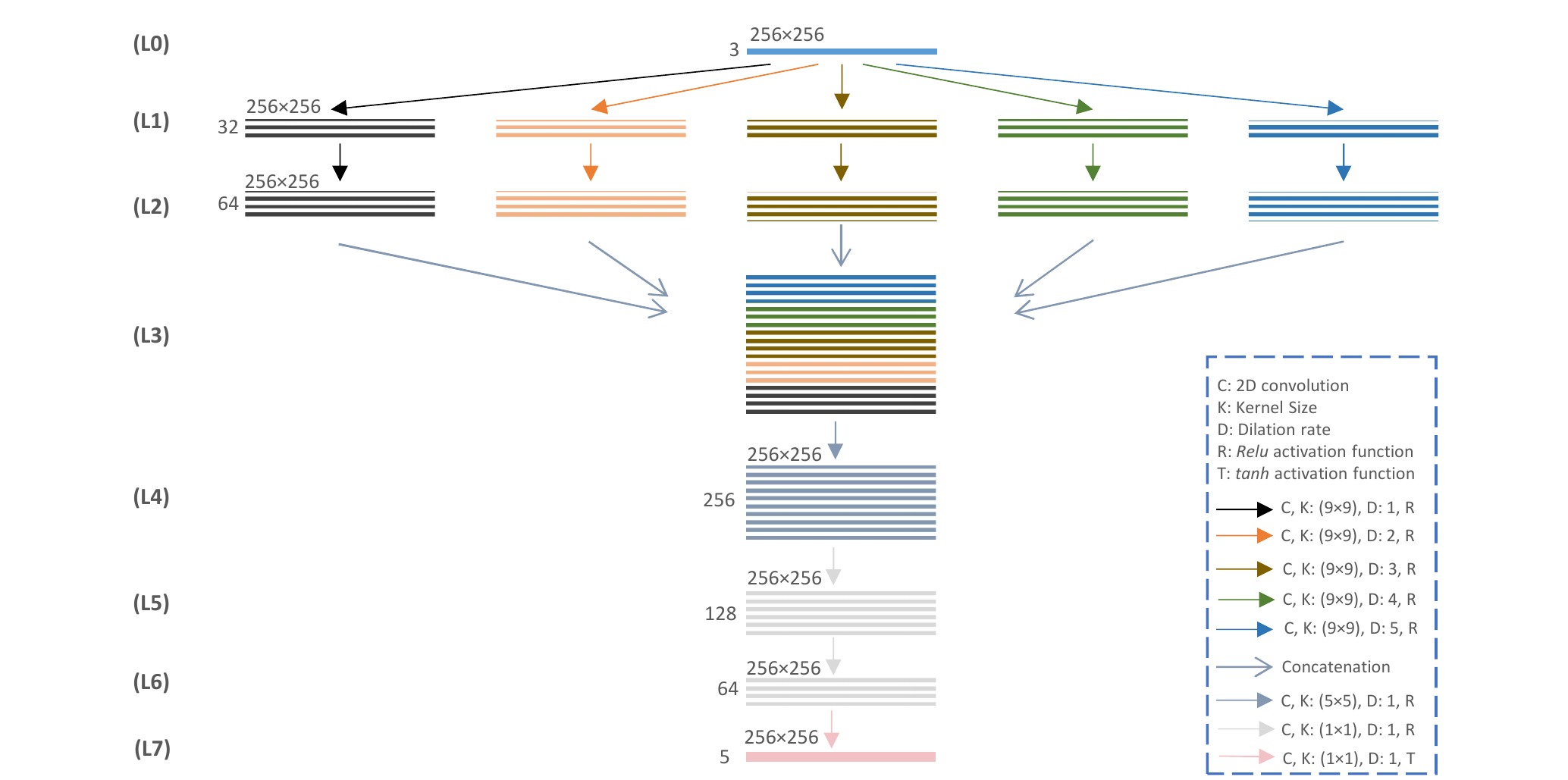}
  \caption{(Color online) Architecture of one of the \textit{Flat-net} networks producing 5 \textit{heat-maps} corresponding to 5 landmarks.}
  \label{fig:Model}
\end{figure*}

The proposed architecture \textit{Flat-net} is presented in Fig. \ref{fig:Model}. As indicated by its name, it does not contain any pooling, down and up sampling nor fully connected layers but considers various kernel sizes and dilation rates.
It is designed to explore in the first layers the image at different resolutions, so as to deal with the different morphologies, and combine the resulting feature maps in a second step to output the desired \textit{heat-maps}.
In layers L1 and L2, convolutional filters of kernel size $9\times9$ are applied consecutively using 5 different dilation rates. The generated feature maps are then concatenated in layer L3. It is followed by a convolutional layer of kernel size $5\times5$ and three consecutive convolutional layers of kernel size $1\times1$. The activation functions for all convolutional layers are \textit{Relu}, except the last layer (L7) using \textit{tanh}. 
The number of filters used for each layer is indicated in the Fig. \ref{fig:Model}. 
The loss is measured as the mean absolute error between the predicted and the desired \textit{heat-maps}.
Since there is no pooling in this architecture, the concatenation of the layer L3 leads to a very large size tensor, causing practical memory issues.
To solve this problem, similarly to the \textit{p2p-GN} approach, the prediction of the 21 \textit{heat-maps} are split in practice into 5 different networks.
Again, the additional number of weights resulting from the use of 5 networks in this approach is discussed in section \ref{sec-results}.

\subsection{Implementation and Evaluation}

The $9\times62$ input grayscale images are converted into grayscale RGB by simple channel repetition to comply with the input format of the networks.
Considering the relatively limited number of data, the dataset is augmented \cite{augmentor} via 10 different arbitrary methods as follows: (1) Addition of a Gaussian intensity noise of mean equal to 0 and variance to $12.75$, (2) Blurring with a Gaussian of variance $\sigma$=5 pixels, (3-4) Rotation of +10 and -5 degrees, (5-6) Translation of (+30,+10) pixels and (+40,-10) pixels, (7) Rotation of -5 degrees followed translation of (+30,+10) pixels, (8) Zooming out of scale 0.8, (9) Translation of (+30,+10) pixels followed by zooming in of scale 1.2, and (10) Translation of (+40,+20) pixels followed by zooming out of scale 0.9 plus blurring with a Gaussian of variance $\sigma$=3 pixels.
By this method, the dataset is artificially augmented from $9\times62=558$ to $11\times9\times62=6138$ images.

The errors of prediction of the landmark coordinates are evaluated by means of Euclidean distance and Root Mean Squared Error (RMSE).
The Euclidean distances are expressed in pixels to comply with the existing results in the literature of the domain while the RMSE is expressed in centimeters to provide comprehensive and interpretable results for speech analyses purposes. In addition, the percentage of samples (i.e. test images) which landmarks presenting distance errors higher than 5 pixels (\textit{i.e.} outliers), are reported.
% \textcolor{blue}{Moreover, statistical significance test via \textit{paired t-test} is used on achieved Euclidean distances to show the significant difference between the results of the proposed method and others.}
Moreover, the statistical significance of the difference between the means of the distance errors obtained for our methods on the one hand and all the other methods on the other hand is evaluated by means of \textit{paired t-test}.

For the $l^{th}$ landmark, if $(x_g,y_g)_l$ and $(x_p,y_p)_l$ denote respectively the ground-truth and the predicted coordinates, the Euclidean distance $d_l$ is calculated as follows:
\begin{gather}
    d_l=\sqrt{(x_g-x_p)_l^{2}+(y_g-y_p)_l^{2}}
\end{gather}
Using similar notations, the RMSE is calculated as follows:
\begin{gather}
RMSE = \sqrt{\frac{1}{Q} \sum_{q=1}^Q (x_g-x_p)_q^{2}+(y_g-y_p)_q^{2} }
\end{gather}
where $Q$ is the number of considered elements, \textit{e.g.} all the landmarks of all images, and $q$ is the corresponding index. 

Most of the landmark localization methods are evaluated in the literature by a hold-out scheme, \textit{i.e.} by splitting the data into train/test sets, for instance $21,997/1,000$ for the \textit{HyperFace} method \cite{HyperFace}, $2,000/330$ for the \textit{dlib} method \cite{dlib} or $3148/600$ for the \textit{DAN} method \cite{DAN}.
In the current application, the ultimate objective is however to localize landmarks on new speakers, \textit{i.e.} on speakers where no data were available before. To comply both with the literature benchmarks and the specificity of the study, the performances are evaluated via two schemes: (1) the randomized 10-fold cross-validation (CV) and (2) the leave-one-subject-out cross-validation (LoSo). 
Note in addition that 5\% of the training data in each training session are randomly set aside in advance for validation purposes and tracking the learning curves.
The learning would stop when the loss curve of the validation set reaches plateau.

In the CV scheme, the augmented dataset is randomly split into 10 groups, 9 being used for training purposes, \textit{i.e.} $5,525$ samples (after rounding), and 1 for test purposes, \textit{i.e.} $613$ samples (after rounding).
The training and evaluation is then repeated 10 times until each single group has been used as the test set. Each sample is therefore being used as a test sample at some point during the process. 
Note that in this evaluation scheme, the train and test sets may share data of same subjects and/or of same articulations, making the two sets not completely independent.
However, in accordance with the literature in the domain, a very large dataset with many more subjects and articulations could be perfectly evaluated through this scheme and is therefore considered in this study.
Above all, this scheme is considered in our evaluation to assess the validity of the hyper-parameters reported by the methods for our environment.

In the LoSo scheme, the 62 images of one arbitrary subject are set aside to serve as the test set.
The remaining images are then augmented, leading to  $8\times62\times11=5,456$ images, and used for the training.
In other words, the network is not trained with data from the test subject.
The training and evaluation is then repeated 9 times until each subject has been used as the test set. Each sample is therefore being used as a test sample at some point during the process on a model trained on the other subjects. 
This scheme is much stricter and challenging that the CV scheme as the trained network does not contain any information regarding the tested subject.
Note however that the train and test sets may still share data of same articulations (but not speakers). This point will be revisited in the discussion.

% Finally, the statistical significance of the difference between the means of the distance errors obtained for our methods on the one hand and all the other methods on the other hand is evaluated by means of paired t-tests. 

The results for the two evaluation schemes are presented in the section \ref{sec-results}.
All of the hyper-parameters of the methods taken from the literature are set to their default values mentioned in their corresponding studies.
The training machine was made of an Intel Xeon w-2145 (3.70 GHz) CPU and a NVIDIA Tesla P100-SXM2-16GB GPU. Except for \textit{dlib}, all the methods are trained on GPU.
All the implementation are available online on GitHub (\href{https://github.com/mohaEs}{https://github.com/mohaEs} and \href{https://github.com/christianpayer/MedicalDataAugmentationTool/tree/master/bin/experiments/localization/hand_xray}{https://github.com/christianpayer/MedicalDataAugmentationTool}).

% #########################
% #########################
% #########################
% #########################
% #########################
% #########################
% #########################
% #########################

\section{Results}
\label{sec-results}

\begin{figure*}[t]
\centering
\captionsetup{justification=centering}
  \includegraphics[width=0.5\linewidth]{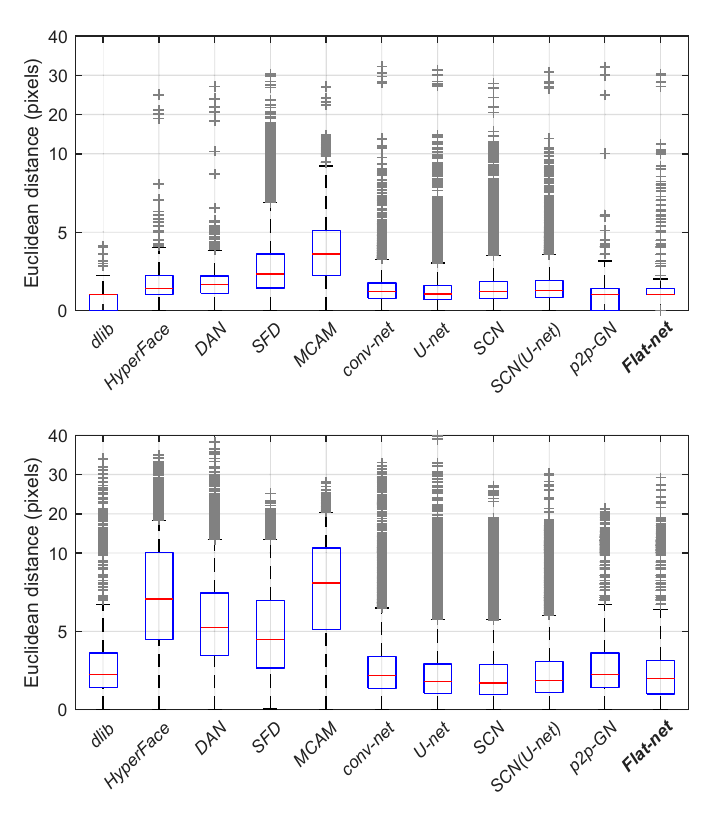}
  % \caption{Box plots of the achieved distance errors between the locations of the predicted and truth keypoints in 10-fold cross-validation scheme (CV).}
  \caption{Box plots of the overall Euclidean distances between the predicted and true landmark locations for the 11 methods considered in the results, for the CV (top) and LoSo (bottom) evaluation schemes. Note the y-axis scale, linear in two steps to zoom in the 0-10 pixels range.}
  \label{fig:boxplots}
\end{figure*}

As noted in the section \ref{sec-methods}, one of the 12 methods considered in this study, the \textit{DS-net}, did not converge. For this reason, unless explicitly mentioned, the results cover only the remaining 11 methods.
An overall comparison of the performances of the eleven methods are provided in Fig. \ref{fig:boxplots} for both the CV and the LoSo schemes. It displays in box plots the Euclidean distances between the predicted and true landmark locations.
% Notice that, the \textit{DS-net} is not able to converge in our problem, so its results are not included in this section. 
For each box, the central mark indicates the median while the bottom and top edges indicate respectively the $25^{th}$ and $75^{th}$ percentiles. The whiskers extend to the most extreme data points not considered as outliers, the outliers being plotted individually using the `$+$' symbol.

Regarding the CV scheme, all the methods show good accuracy, with boxes below 2.5 pixels, except the \textit{MCAM} method. This might be ascribed to the design of the method, optimized for joint localization and not anatomical landmarks as in the current case.
Nevertheless, the method still shows decent results, with $75\%$ of the distance errors being smaller than 5 pixels.
On the other side, the \textit{dlib} method presents the best results, possibly due to its boosting approach.

By attempting to predict landmarks on a subject not used to train the models, the LoSo evaluation scheme is more constraining and presents logically deteriorated -- but more pertinent -- results in comparison to the CV evaluation scheme.
% The results for the \textit{DAN}, \textit{MCAM}, \textit{SFD} and \textit{HyperFace} methods appear in particular significantly deteriorated.
The results for the \textit{HyperFace}, \textit{DAN}, \textit{SFD} and \textit{MCAM} methods appear in particular significantly deteriorated.
On the contrary, the deterioration appears more limited for the other methods and still lead to fairly good accuracy, with boxes remaining below 3.5 pixels. 

% The best results are achieved for the \textit{Flat-net} architecture, while the \textit{p2p-GN} architecture and the \textit{dlib} method still lead to fairly good accuracy, with boxes remaining below 3.5 pixels.

\begin{figure*}[t]
\centering
\captionsetup{justification=centering}
\includegraphics[width=\linewidth]{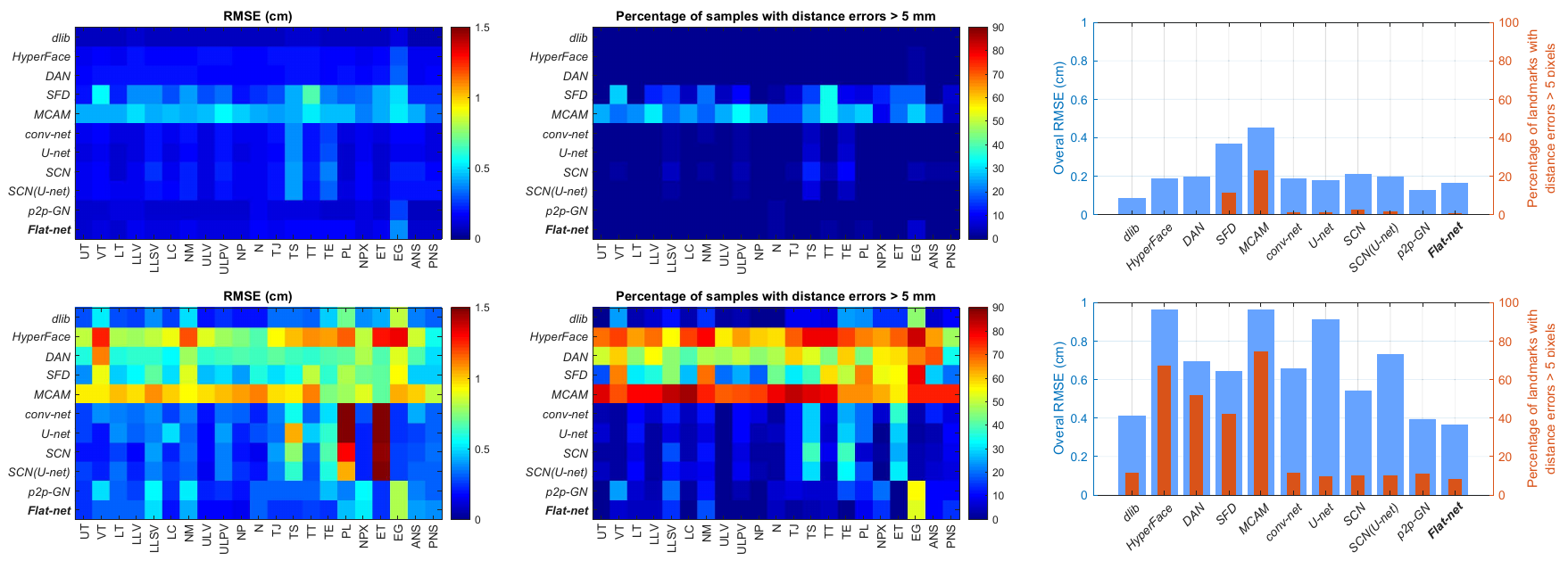}
  \caption{(Color online) RMSE and percentage of samples with errors higher than 5 mm for the 11 methods considered in the results  for the CV scheme (top) and the LoSo scheme (bottom). \\ \footnotesize{(Left) RMSE matrix for each landmark. (Middle) Matrix of the percentage of samples with errors > 5 mm for each landmark. (Right) Overall RMSEs and percentages.}}
  \label{fig:comp}
\end{figure*}

The results in terms of RMSE and per landmark are provided in Fig. \ref{fig:comp}. 
%The figure shows that the worst results are achieved for the EG, NM and VT landmarks. 
It confirms the lower accuracy already noted for the LoSo scheme in comparison to the CV scheme. 
It also shows that the four methods \textit{HyperFace}, \textit{DAN}, \textit{SFD} and \textit{MCAM} estimate in the LoSo scheme many landmarks for more than 50\% of the images with an error larger than 0.5~cm (5 pixels), rather problematic for speech production studies. 
% The \textit{conv-net}, \textit{U-net}, \textit{SCN} and \textit{SCN(U-net)} methods has a good results generally, they have serious error on PL and ET landmarks.
The \textit{conv-net}, \textit{U-net}, \textit{SCN} and \textit{SCN(U-net)} methods display decent results in general but present serious errors on the PL and ET landmarks.
On the contrary, the three methods \textit{dlib}, \textit{p2p-GN} and \textit{Flat-net} still show acceptable results, with almost all landmarks for more than 70\% of the images having an error lower than 0.5~cm. For these methods, the landmark EG is the most challenging one.
% It means that the three some methods are not suitable to handle satisfactorily the data and problem presented in this study. A larger dataset with significantly more subjects and articulations might solve this issue. On the contrary, the three latter methods provide fairly good results in the LoSo scheme despite the limited dataset.
It means that our method, the \textit{Flat-net}, performs at least as good as the two best methods taken from the literature, namely the \textit{dlib} and \textit{p2p-GN} methods. They provide fairly good results in the LoSo scheme despite the limited dataset. The other methods do not appear suitable to handle satisfactorily the data and problem presented in this study. A larger dataset with significantly more subjects and articulations might solve this issue.

A summary of the results and characteristics of the twelve methods of the study is reported in Table \ref{tab:comp}.
Note that the storage space is reported for information but does not play in our eyes a critical role. Moreover, it depends on the data format and of the extent of meta-data stored. Also training time depends on the learning rate, type of optimizer and used deep learning framework (Pytorh vs. Tensorflow, etc). 
In overall, the best results in LoSo scheme are reported for our method, the \textit{Flat-net} method, with a RMSE of 0.36~cm, slightly better than for the \textit{dlib} (0.39) and \textit{p2p-GN} (0.41) methods. 
Among all methods, the DL-based methods require significantly more training time and storage space. The \textit{p2p-GN} method in particular has the largest number of weights to train and requires the largest space for storage. The \textit{dlib} method is on the contrary rapidly trained, even on CPU, and requires the smallest space for storage.
% \textcolor{blue}{Table \ref{tab:p-values} shows the results of statistical test and p-values of two sample t-test between the achieved distance errors by \textit{Flat-net} and other methods. Table \ref{tab:p-values} confirms that there is a significant difference between results of \textit{Flat-net} and others with all lower than 0.05. }

% Given that (1) our objective is to compare the performance of our method in reference to other existing methods of the literature and (2) our method presents in general the highest accuracy, paired t-tests have been conducted between the distance errors observed for the \textit{Flat-net} method on the one hand and for the remaining 10 methods on the other hand.
The results of the t-tests conducted between the distance errors obtained by our method, the \textit{Flat-net} method, also presenting interestingly the best performance, and the errors obtained by the other methods are presented in Table \ref{tab:p-values}.
% It tests whether the gap in performance observed in Table \ref{tab:comp} is statically significant or not.
% The results are provided in Table \ref{tab:p-values} and confirm that there is a significant difference between the \textit{Flat-net} and the other methods (p<.05), reinforcing our previous observations.
It confirms a statistical significance between the means of the distance errors, all of them obtained with p=0.000 except p<0.04 for the \textit{SCN} method.

\begin{table}[h]
% \caption{Summary of the results and characteristics of the twelve methods of the study. The lowest occurrence for each line is written in blue. \footnotesize{The sign '-' means the information is not relevant or available. \textit{DS-net} method did not converge in our study.  }}
\caption{Summary of the results and characteristics of the twelve methods of the study. The lowest occurrence for the four left columns  is emphasized in blue. \footnotesize{The sign '-' means that the information is not relevant or not available. Remember that the \textit{DS-net} method did not converge in our study.  }}
\label{tab:comp}
\begin{tabular}{|c|c|c|c|c|c|c|c|c|}
\hline
\textbf{}                    & \textbf{\begin{tabular}[c]{@{}c@{}}RMSE\\ in CV \\ (cm)\end{tabular}} & \textbf{\begin{tabular}[c]{@{}c@{}}RMSE\\ in LoSo \\ (cm)\end{tabular}} & \textbf{\begin{tabular}[c]{@{}c@{}}Distance error \\ in CV \\ (pixels)\end{tabular}} & \textbf{\begin{tabular}[c]{@{}c@{}}Distance error \\ in LoSo\\  (pixels)\end{tabular}} & \textbf{\begin{tabular}[c]{@{}c@{}}Number\\  of \\ weights\end{tabular}} & \textbf{\begin{tabular}[c]{@{}c@{}}Number \\ of \\ epochs\end{tabular}} & \textbf{\begin{tabular}[c]{@{}c@{}}Training\\ time\\ (min)\end{tabular}} & \textbf{\begin{tabular}[c]{@{}c@{}}Storage \\ space\\ (MB)\end{tabular}} \\ \hline
\textit{\textbf{dlib}}       & \color{blue} 0.08   & 0.41  & \color{blue} 0.59 $\pm$ 0.59 &  2.94 $\pm$ 2.74  & -                 & -         & 55 (cpu)         & 45      \\ \hline
\textit{\textbf{HyperFace}}  & 0.18   & 0.96  &   1.49 $\pm$ 0.99   &  8.02 $\pm$ 4.91          & 35,168,006       & 200        & 50                 & 402           \\ \hline
\textit{\textbf{DAN}}        & 0.19   & 0.69  &   1.66 $\pm$ 0.91   &  5.81 $\pm$ 3.53          & 2 $\times$ 23,104,092  & 2 $\times$ 120   & 2$\times$130   & 2$\times$280 \\ \hline
\textit{\textbf{SFD}}        & 0.36   & 0.64  &  2.85 $\pm$ 2.21    &      5.23 $\pm$ 3.51       &    6,248,029      &     150     &    -     &     25      \\ \hline
\textit{\textbf{MCAM}}       & 0.45   & 0.96  &  3.96 $\pm$ 1.96    &   8.47 $\pm$ 4.18         & 14,500,480       & 60         & 150      & 210     \\ \hline
\textit{\textbf{DS-net}}     & -  & - & -  &    -  &      2,056,741         &     -       &       -        &       16        \\ \hline
\textit{\textbf{conv-net}}   & 0.18   & 0.65  & 1.39 $\pm$ 1.14  &    3.34 $\pm$ 5.47  &   9,933,349   &   40    &    117      &   78      \\ \hline
\textit{\textbf{U-net}}      & 0.17   & 0.91  &  1.29 $\pm$ 1.17    &   3.07 $\pm$ 8.38  &   2,662,181   &     80   &    198   &     21  \\ \hline
\textit{\textbf{SCN}}        & 0.20   & 0.54  &  1.52 $\pm$ 1.38    &   2.66 $\pm$ 4.61     &    1,240,958         &      100      &     225       &      10        \\ \hline
\textit{\textbf{SCN(U-net)}} & 0.19   & 0.73  &  1.51 $\pm$ 1.21   &   2.92 $\pm$ 6.55   &     3,718,602     &    60    &  175  &   30     \\ \hline
\textit{\textbf{p2p-GN}}     & 0.12   & 0.39  &  0.89 $\pm$ 0.85    &   2.93 $\pm$ 2.51           & 7$\times$54,420,483  & 7$\times$60   & 7$\times$55  & 7$\times$440  \\ \hline
\textit{\textbf{Flat-net}}   & 0.16   & \color{blue} 0.36  &  1.12 $\pm$ 1.13 &  \color{blue} 2.39 $\pm$ 2.47 & 5$\times$2,958,533  & 5$\times$30    & 5$\times$88  & 5$\times$34    \\ \hline
\end{tabular}
\end{table}

\begin{table}[]
\centering
% \caption{p-value of two sample t-test for distance errors between \textit{Flat-net} and other methods.}
\caption{p-values of paired t-tests for distance errors between the \textit{Flat-net} and the remaining 10 methods considered in the results for the CV and LoSo schemes.}
\begin{tabular}{|c|c|c|c|c|c|c|c|c|c|c|}
\hline
 & \textit{\textbf{dlib}} & \textit{\textbf{HyperFace}} & \textit{\textbf{DAN}} & \textit{\textbf{SFD}} & \textit{\textbf{MCAM}} & \textit{\textbf{Conv-net}} & \textit{\textbf{U-net}} & \textit{\textbf{SCN}} & \textit{\textbf{SCN (U-net)}} & \textit{\textbf{p2p-GN}} \\ \hline
CV &      0   &    2.2e-150      &  0   &    0     &   0  &   2.1e-72    &  1.7e-29     &  3.6e-125   &  5.3e-137       &    5.5e-70             \\ \hline
LoSo &   9.1e-28      &     0     &    0    &    0     &  0  &    5.3e-44   &  3.9e-10     &   0.037    &        4.6e-08  &      1.7e-28           \\ \hline
\end{tabular}
\label{tab:p-values}
\end{table}

\begin{figure}[h]
\centering
\captionsetup{justification=centering}
  \includegraphics[width=0.5\linewidth,trim={3.5cm 10cm 3.5cm 10cm},clip]{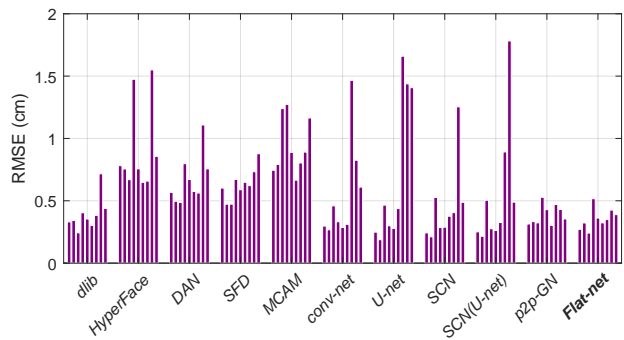}
  \caption{Bar plots of the overall RMSEs per tested subject in LoSo scheme for the 11 methods considered in the results.}
  \label{fig:bar-speaker}
\end{figure}

\begin{figure*}[h]
\centering
\captionsetup{justification=centering}
\includegraphics[width=0.91\linewidth]{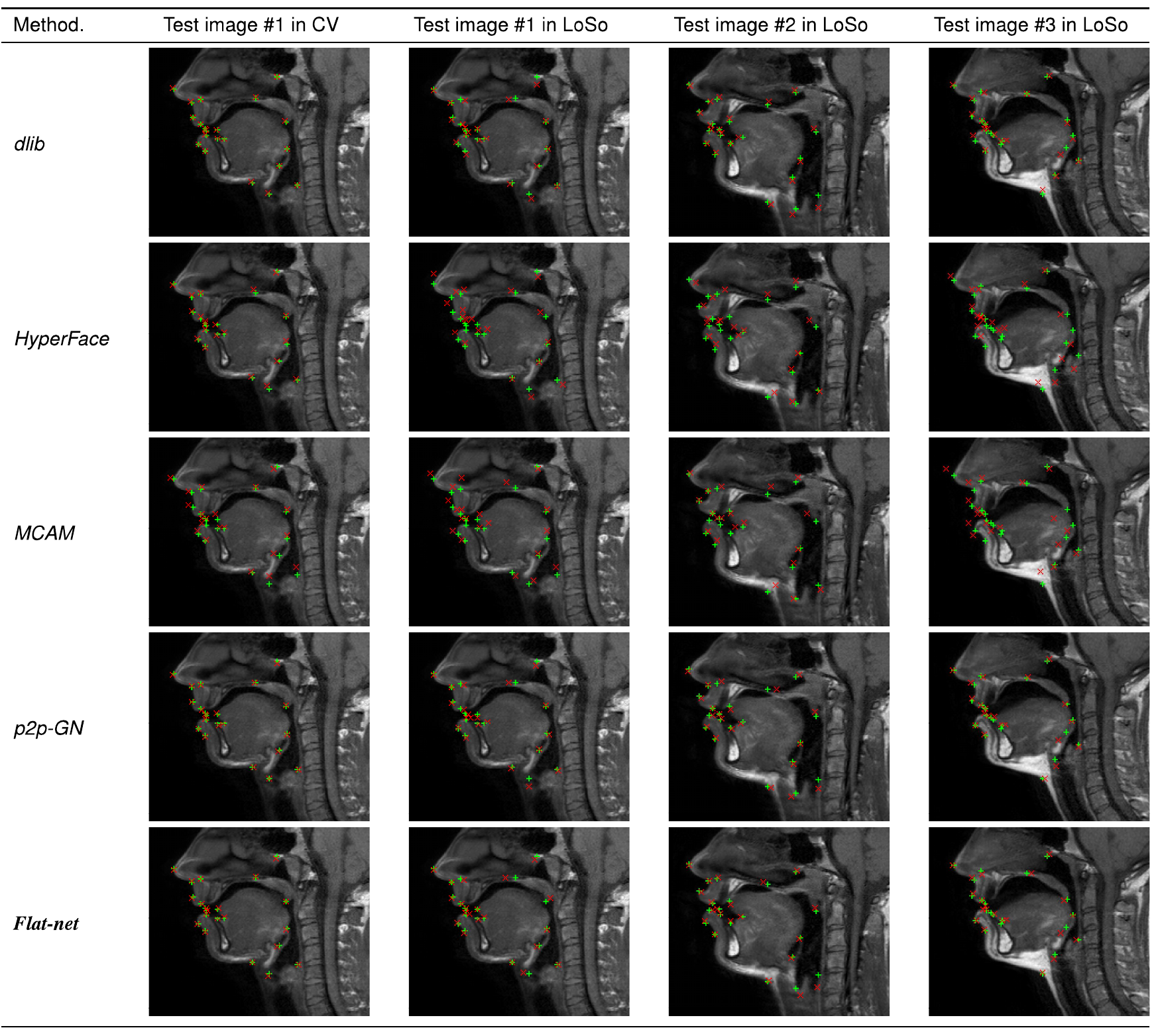}
  \caption{(Color online) MRI images of three different subjects and articulations zoomed in the region of the vocal tract and superimposed with the 21 ground truth (green crosses) and predicted (red crosses) landmarks for the best three methods (\textit{dlib}, \textit{p2p-GN} and \textit{Flat-net}) and the two worst methods (\textit{HyperFace} and \textit{MCAM}) reported for the LoSo scheme. The first test image is also presented with results of the CV scheme.}
  \label{tab:shows}
\end{figure*}

The overall RMSEs per subject in LoSo scheme for the eleven methods considered in the results are displayed in Fig. \ref{fig:bar-speaker}.
The \textit{p2p-GN} and \textit{Flat-net} methods tend to show more homogeneous and lower errors across subjects, suggesting that they are the two more accurate and robust methods to predict the landmarks for new and unseen subjects.
Finally, some examples of practical results for the three best and the two worst methods are illustrated in Figure \ref{tab:shows}.

\section{Discussion}
\label{sec-discussion}

% The present paper described twelve methods to localize anatomical landmarks of the vocal tract area on MRI images.
The present paper described an original method to localize anatomical landmarks of the vocal tract area on MRI images and compared its performance to 11 state-of-the-art methods of the literature.
A dataset of midsagittal MRI from 9 speakers sustaining 62 articulations and annotated with the location of 21 landmarks has been considered.
% Twelve different methods are investigated for this purpose.
The methods have been evaluated through two schemes, a randomized 10-fold scheme and a leave-one-speaker-out scheme, considered as more challenging.
Experimental results show the ability of all methods to cope with the problem in the 10-fold scheme but divergence of performance appear in the more challenging leave-one-speaker-out scheme. 
% In general, the best results are achieved for the three methods \textit{dlib}, \textit{p2p-GN} and \textit{Flat-net}.
In general, our method, the \textit{Flat-net} method, outperforms the other methods to solve the specific problem of the study and is only approached by two other methods, namely \textit{dlib} and \textit{p2p-GN}.
Interestingly, these two methods include the only method not based on DL (\textit{dlib}) as well as the method that has been adapted the most to fit at best our problem (\textit{p2p-GN}).
% two methods specifically designed to solve the current problem.
% Also, methods based on heat-map regression are seriously practical and better than others.
Note also that the \textit{heat-map}-based methods present in general significantly better results than the other methods, supporting this approach to tackle the problem of landmark localization on medical images.
This approach leads to networks without fully connected layers, usually used to transform the feature maps into vectors of landmark locations in output. This may result in an architecture possibly less prone to error propagation, especially for such a limited dataset.
% Indeed, although heat-maps are also somehow part of the \textit{DAN} and \textit{MCAM} methods, the design in channels as well as the use of networks able to output directly the heat-maps are successful and methods based on heat-map regression bring more accurate results.
Indeed, although \textit{heat-maps} are also somehow part of the \textit{DAN} and \textit{MCAM} methods, the design in channels together with the use of adapted networks able to output directly these \textit{heat-maps} proved successful for all the other methods.

The good performance of the \textit{dlib} method might be ascribed to the use of the boosting approach combined with the analysis of the input image in small regions by applying windows, possibly reducing the sensitivity to the variability of other regions of the image.
% The good performance of the \textit{p2p-GN} and \textit{Flat-net} methods might be ascribed to the combination of the use of \textit{heat-maps in channels} together with adapted networks.
In overall, our \textit{Flat-net} method, the only method entirely developed in this study, tends to be more robust and to present better performances.
One could object that the better performance of the \textit{Flat-net} network may simply come from its hyper-parameters optimized for the current problem. The results obtained in the CV evaluation scheme suggest however that all eleven methods considered in the results -- except the \textit{MCAM} method to a certain extent -- can perform the task with success, discarding this objection.
These results support the approach using \textit{heat-maps in channels} and networks without fully connects layers to localize landmarks of the vocal tract area on MRI data.

\begin{figure}[ht]
\centering
\captionsetup{justification=centering}
  \includegraphics[width=0.5\linewidth,trim={3.5cm 10cm 3.5cm 10cm},clip]{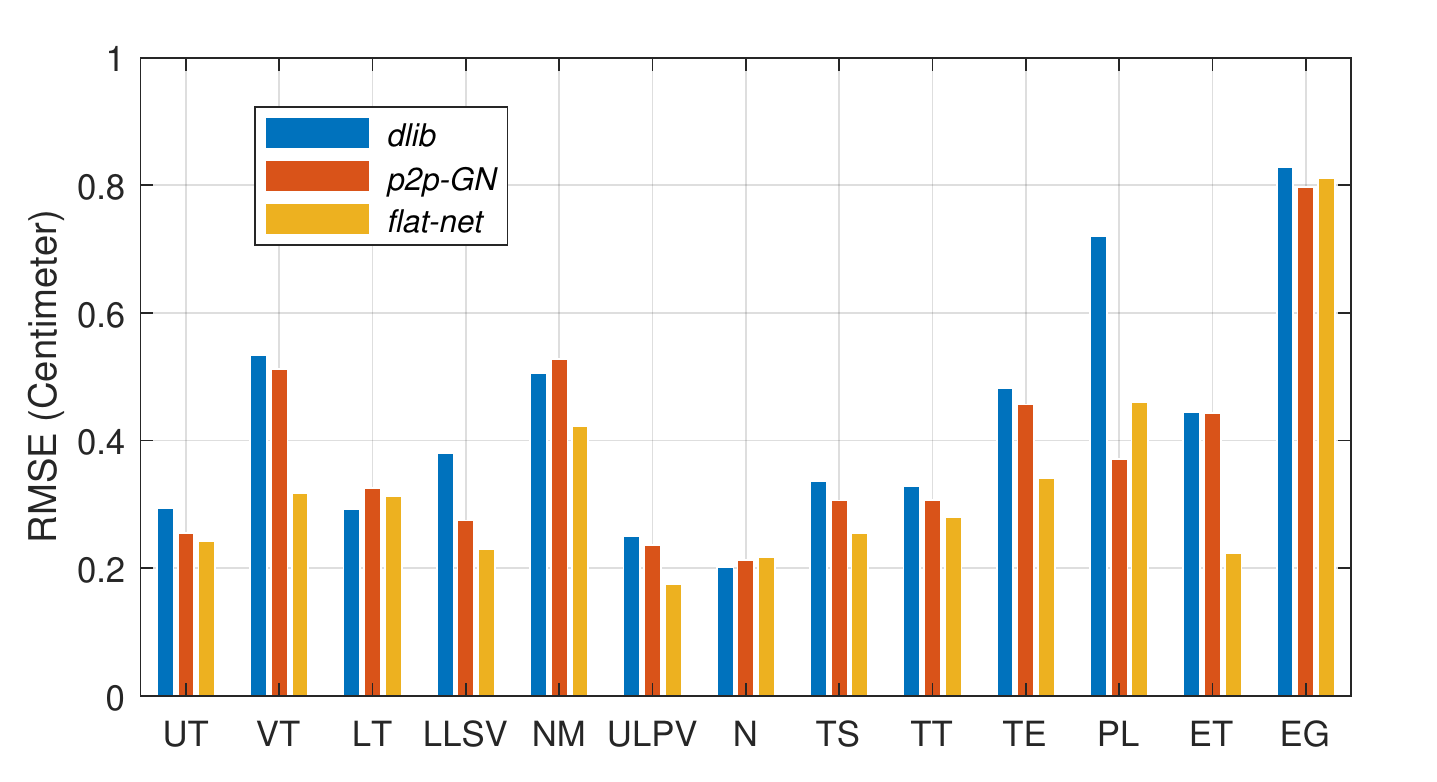}

  \caption{(Color online) Bar plots of the overall RMSEs in LoSo scheme for the 3 methods \textit{dlib}, \textit{p2p-GN} and \textit{Flat-net} for a subset of 13 out of 21 landmarks.}
  \label{fig:bar-some}
\end{figure}

Regarding the localization of the landmarks, the overall RMSEs in LoSo scheme for the three best methods identified above and for a subset of 13 out of 21 landmarks are displayed in Fig. \ref{fig:bar-some}: the 11 landmarks mentioned in the section \ref{subsec-challenges} plus the landmarks N on the one hand and NM and EG on the other hand, presenting respectively the best and worst results. Note that these results are a zoom of the results presented in Fig. \ref{fig:comp}a.
All these landmarks are more accurately located by the \textit{Flat-net} method, occasionally at the same precision than other methods. Except for NM, PL and EG, the RMSEs stay below 0.35~cm, a fairly good result considering the challenge associated with these landmarks and comparable to the overall RMSE achieved by the method. The landmarks UT and LT in particular, not visible on MRI and giving many problems in articulatory speech studies \cite{takemoto2004}, are estimated with a respective accuracy of 0.25~cm and 0.3~cm for new speakers without additional \textit{a priori} information. Similarly, the landmark TS, important for speech articulations and challenging to identify \cite{ananthakrishnan2010}, is estimated with a precision of 0.25~cm. This emphasizes the robustness of the chosen approach through deep learning and the potential of these results for speech studies. The worst results are achieved for the landmarks NM, outside the range of the vocal tract area, and EG and PL, in the larynx area. Although the larynx plays an important role in speech production, it remains at the margin of the vocal tract area and does not play a central role in articulatory speech studies \cite{s_serrurier2019b}.

It should also be noted that some landmarks do not exhibit very salient characteristics, such as the junctions between the wet and dry vermillion of the lips (ULV and LLV) or the sublingual cavity posterior point (TS) when the tip of the tongue is in a low position. Similarly, some regions tend to be hardly interpretable in terms of anatomy, such the anterior part of the larynx associated with the landmark EG. Annotating manually their exact location on the images was a challenging task at the first point, questioning the quality of the ground-truth data and possibly explaining the lower accuracy achieved for EG. Furthermore, a detailed analysis of the results such as presented in Figure \ref{tab:shows} reveals that the location of some landmarks may appear occasionally more accurate in output of the presented methods than in the original so-called ground-truth. This is a well known effect of machine learning methods, and DL methods in particular, which tend to avoid encoding noise and outliers by means of regularization techniques \cite{regularization, book}. In general, a larger dataset labelled by several experts may limit the impact of the uncertainties in the ground-truth data and reinforce the robustness the DL methods and their resistance to noise.

The methods have been evaluated by means of two schemes, the CV  and LoSo schemes.
Strictly speaking, the most rigorous scheme would have been to leave both subject and articulation out, \textit{i.e.} leave-one-subject-one-class-out (LoSoCo), to ensure that the network does not contain any information regarding the new tested image. In this scheme, all the articulations of one subject and one specific articulation for all subjects would be discarded in the training and the same specific articulation of the left subject would be tested. This would lead to $62 \times 9 = 558$ training sessions for each of the twelve methods. According to the times reported in Table \ref{tab:comp}, it would take more than a year, making this evaluation unrealistic in practice. However, the challenge of the problem lies rather in the estimation of the landmark locations for a new speaker than for a new articulation. Indeed, the corpus of 62 articulations can be considered large enough and representative of the French phonemic repertoire so that one articulation could fairly well be estimated from the 61 others \cite{s_serrurier2019b}. For this reason, the LoSo scheme appears as a valid approximation to evaluate the methods on our problem.

In summary, the method proposed in this study (\textit{Flat-net}) outperforms the state-of-the-art methods. It supports the description of landmarks locations in terms of \textit{heat-maps in channels} and the generation of these \textit{heat-maps} by means of DL networks without fully connected layers for such a variable and limited dataset.
Future works may include the combination of successful features from the \textit{dlib} method, showing very promising results, with DL approaches to create more robust methods, such as for instance the methods using deep forest networks \cite{deepforest, ensemble}.
% In addition, since the accuracy of the heat-map-based methods for all landmarks are not same, it is reasonable to make assembled machine via combining them and ensemble machine learning.
In addition, since the accuracy of the \textit{heat-map}-based methods for all landmarks are not same, it seems reasonable to combine them to form so-called machine learning ensemble methods.
% For example, \textit{p2p-GN} and \textit{Flat-net} are good for landmarks ET, PL, TE and TS but not for EG and NPX and on contrary SCN is good on EG and NPX but not on ET, PL, TE and TS.
For example, the \textit{p2p-GN} and \textit{Flat-net} methods present good accuracy for the landmarks ET, PL, TE and TS but not for the landmarks EG and NPX while on the contrary the \textit{SCN} method presents good accuracy for the landmarks EG and NPX but not for the landmarks ET, PL, TE and TS.
% This means, it may possible to find a new network architecture or combination method to fuse the results and exploit the advantages of different methods.
Combining these methods to take advantage of their relative assets may therefore lead to higher accuracy in the localization of vocal tract area landmarks from MRI data together with more robustness.
Furthermore, considering the recent rise of real-time MRI for speech production studies \cite{review-vt-apps}, the next steps will be to adapt this technique to real-time MRI data.

% #########################
% #########################
% #########################
% #########################
% #########################
% #########################
% #########################
% #########################

% #########################
% #########################
% #########################
% #########################
% #########################
% #########################
% #########################
% #########################

\bibliography{refs}

\begin{thebibliography}{10}
\urlstyle{rm}
\expandafter\ifx\csname url\endcsname\relax
  \def\url#1{\texttt{#1}}\fi
\expandafter\ifx\csname urlprefix\endcsname\relax\def\urlprefix{URL }\fi
\expandafter\ifx\csname doiprefix\endcsname\relax\def\doiprefix{DOI: }\fi
\providecommand{\bibinfo}[2]{#2}
\providecommand{\eprint}[2][]{\url{#2}}

\bibitem{harshman1977}
\bibinfo{author}{Harshman, R.}, \bibinfo{author}{Ladefoged, P.} \&
  \bibinfo{author}{Goldstein, L.}
\newblock \bibinfo{journal}{\bibinfo{title}{Factor analysis of tongue shapes}}.
\newblock {\emph{\JournalTitle{The Journal of the Acoustical Society of
  America}}} \textbf{\bibinfo{volume}{62}}, \bibinfo{pages}{693--707}
  (\bibinfo{year}{1977}).

\bibitem{beautemps2001}
\bibinfo{author}{Beautemps, D.}, \bibinfo{author}{Badin, P.} \&
  \bibinfo{author}{Bailly, G.}
\newblock \bibinfo{journal}{\bibinfo{title}{Linear degrees of freedom in speech
  production: Analysis of cineradio- and labio-film data and
  articulatory-acoustic modeling}}.
\newblock {\emph{\JournalTitle{The Journal of the Acoustical Society of
  America}}} \textbf{\bibinfo{volume}{109}}, \bibinfo{pages}{2165--2180}
  (\bibinfo{year}{2001}).

\bibitem{s_serrurier2019b}
\bibinfo{author}{Serrurier, A.}, \bibinfo{author}{Badin, P.},
  \bibinfo{author}{Lamalle, L.} \& \bibinfo{author}{Neuschaefer-Rube, C.}
\newblock \bibinfo{journal}{\bibinfo{title}{{Characterization of inter-speaker
  articulatory variability: a two-level multi-speaker modelling approach based
  on MRI data}}}.
\newblock {\emph{\JournalTitle{The Journal of the Acoustical Society of
  America}}} \textbf{\bibinfo{volume}{145}}, \bibinfo{pages}{2149--2170},
  \doiprefix\url{10.1121/1.5096631} (\bibinfo{year}{2019}).

\bibitem{yamasaki2017vocal}
\bibinfo{author}{Yamasaki, R.} \emph{et~al.}
\newblock \bibinfo{journal}{\bibinfo{title}{Vocal tract adjustments of
  dysphonic and non-dysphonic women pre-and post-flexible resonance tube in
  water exercise: a quantitative mri study}}.
\newblock {\emph{\JournalTitle{Journal of Voice}}}
  \textbf{\bibinfo{volume}{31}}, \bibinfo{pages}{442--454}
  (\bibinfo{year}{2017}).

\bibitem{guzman2017computerized}
\bibinfo{author}{Guzman, M.} \emph{et~al.}
\newblock \bibinfo{journal}{\bibinfo{title}{Computerized tomography measures
  during and after artificial lengthening of the vocal tract in subjects with
  voice disorders}}.
\newblock {\emph{\JournalTitle{Journal of voice}}}
  \textbf{\bibinfo{volume}{31}}, \bibinfo{pages}{124--e1}
  (\bibinfo{year}{2017}).

\bibitem{VPI}
\bibinfo{author}{Freitas, A.~C.}, \bibinfo{author}{Wylezinska, M.},
  \bibinfo{author}{Birch, M.~J.}, \bibinfo{author}{Petersen, S.~E.} \&
  \bibinfo{author}{Miquel, M.~E.}
\newblock \bibinfo{journal}{\bibinfo{title}{Comparison of cartesian and
  non-cartesian real-time mri sequences at 1.5 t to assess velar motion and
  velopharyngeal closure during speech}}.
\newblock {\emph{\JournalTitle{PloS one}}} \textbf{\bibinfo{volume}{11}},
  \bibinfo{pages}{e0153322} (\bibinfo{year}{2016}).

\bibitem{disorders}
\bibinfo{author}{De~Alarc{\`o}n, A.}, \bibinfo{author}{Prager, J.},
  \bibinfo{author}{Rutter, M.} \& \bibinfo{author}{Wootten, C.~T.}
\newblock \bibinfo{title}{Laryngeal and tracheal airway disorders}.
\newblock In \emph{\bibinfo{booktitle}{Kendig's Disorders of the Respiratory
  Tract in Children}}, \bibinfo{pages}{1118--1124}
  (\bibinfo{publisher}{Elsevier}, \bibinfo{year}{2019}).

\bibitem{MriSeg}
\bibinfo{author}{Labrunie, M.} \emph{et~al.}
\newblock \bibinfo{journal}{\bibinfo{title}{Automatic segmentation of speech
  articulators from real-time midsagittal mri based on supervised learning}}.
\newblock {\emph{\JournalTitle{Speech Communication}}}
  \textbf{\bibinfo{volume}{99}}, \bibinfo{pages}{27--46}
  (\bibinfo{year}{2018}).

\bibitem{narayanan2004}
\bibinfo{author}{Narayanan, S.}, \bibinfo{author}{Nayak, K.},
  \bibinfo{author}{Lee, S.}, \bibinfo{author}{Sethy, A.} \&
  \bibinfo{author}{Byrd, D.}
\newblock \bibinfo{journal}{\bibinfo{title}{{{A}n approach to real-time
  magnetic resonance imaging for speech production}}}.
\newblock {\emph{\JournalTitle{The Journal of the Acoustical Society of
  America}}} \textbf{\bibinfo{volume}{115}}, \bibinfo{pages}{1771--1776},
  \doiprefix\url{10.1121/1.1652588} (\bibinfo{year}{2004}).

\bibitem{story2005b}
\bibinfo{author}{Story, B.~H.}
\newblock \bibinfo{journal}{\bibinfo{title}{Synergistic modes of vocal tract
  articulation for american english vowels}}.
\newblock {\emph{\JournalTitle{The Journal of the Acoustical Society of
  America}}} \textbf{\bibinfo{volume}{118}}, \bibinfo{pages}{3834--3859},
  \doiprefix\url{10.1121/1.2118367} (\bibinfo{year}{2005}).
\newblock \eprint{https://doi.org/10.1121/1.2118367}.

\bibitem{echternach2016}
\bibinfo{author}{Echternach, M.}, \bibinfo{author}{Burk, F.},
  \bibinfo{author}{Burdumy, M.}, \bibinfo{author}{Traser, L.} \&
  \bibinfo{author}{Richter, B.}
\newblock \bibinfo{journal}{\bibinfo{title}{Morphometric differences of vocal
  tract articulators in different loudness conditions in singing}}.
\newblock {\emph{\JournalTitle{PLOS ONE}}} \textbf{\bibinfo{volume}{11}},
  \bibinfo{pages}{1--17}, \doiprefix\url{10.1371/journal.pone.0153792}
  (\bibinfo{year}{2016}).

\bibitem{review-vt-apps}
\bibinfo{author}{Ramanarayanan, V.} \emph{et~al.}
\newblock \bibinfo{journal}{\bibinfo{title}{Analysis of speech production
  real-time mri}}.
\newblock {\emph{\JournalTitle{Computer Speech \& Language}}}
  \textbf{\bibinfo{volume}{52}}, \bibinfo{pages}{1--22} (\bibinfo{year}{2018}).

\bibitem{valve}
\bibinfo{author}{Zheng, Y.} \emph{et~al.}
\newblock \bibinfo{journal}{\bibinfo{title}{Automatic aorta segmentation and
  valve landmark detection in c-arm ct for transcatheter aortic valve
  implantation}}.
\newblock {\emph{\JournalTitle{IEEE transactions on medical imaging}}}
  \textbf{\bibinfo{volume}{31}}, \bibinfo{pages}{2307--2321}
  (\bibinfo{year}{2012}).

\bibitem{Cephalometric}
\bibinfo{author}{Lindner, C.} \emph{et~al.}
\newblock \bibinfo{journal}{\bibinfo{title}{Fully automatic system for accurate
  localisation and analysis of cephalometric landmarks in lateral
  cephalograms}}.
\newblock {\emph{\JournalTitle{Scientific reports}}}
  \textbf{\bibinfo{volume}{6}}, \bibinfo{pages}{33581} (\bibinfo{year}{2016}).

\bibitem{geometric}
\bibinfo{author}{Vandaele, R.} \emph{et~al.}
\newblock \bibinfo{journal}{\bibinfo{title}{Landmark detection in 2d bioimages
  for geometric morphometrics: a multi-resolution tree-based approach}}.
\newblock {\emph{\JournalTitle{Scientific reports}}}
  \textbf{\bibinfo{volume}{8}}, \bibinfo{pages}{538} (\bibinfo{year}{2018}).

\bibitem{heatmap_2016}
\bibinfo{author}{Payer, C.}, \bibinfo{author}{{\v{S}}tern, D.},
  \bibinfo{author}{Bischof, H.} \& \bibinfo{author}{Urschler, M.}
\newblock \bibinfo{title}{Regressing heatmaps for multiple landmark
  localization using cnns}.
\newblock In \emph{\bibinfo{booktitle}{International Conference on Medical
  Image Computing and Computer-Assisted Intervention}},
  \bibinfo{pages}{230--238} (\bibinfo{organization}{Springer},
  \bibinfo{year}{2016}).

\bibitem{heatmap_2019}
\bibinfo{author}{Payer, C.}, \bibinfo{author}{{\v{S}}tern, D.},
  \bibinfo{author}{Bischof, H.} \& \bibinfo{author}{Urschler, M.}
\newblock \bibinfo{journal}{\bibinfo{title}{Integrating spatial configuration
  into heatmap regression based cnns for landmark localization}}.
\newblock {\emph{\JournalTitle{Medical Image Analysis}}}
  \textbf{\bibinfo{volume}{54}}, \bibinfo{pages}{207--219}
  (\bibinfo{year}{2019}).

\bibitem{eye}
\bibinfo{author}{De~Zanet, S.~I.} \emph{et~al.}
\newblock \bibinfo{journal}{\bibinfo{title}{Landmark detection for fusion of
  fundus and mri toward a patient-specific multimodal eye model}}.
\newblock {\emph{\JournalTitle{IEEE transactions on biomedical engineering}}}
  \textbf{\bibinfo{volume}{62}}, \bibinfo{pages}{532--540}
  (\bibinfo{year}{2014}).

\bibitem{microscope}
\bibinfo{author}{Wang, C.-W.}, \bibinfo{author}{Ka, S.-M.} \&
  \bibinfo{author}{Chen, A.}
\newblock \bibinfo{journal}{\bibinfo{title}{Robust image registration of
  biological microscopic images}}.
\newblock {\emph{\JournalTitle{Scientific reports}}}
  \textbf{\bibinfo{volume}{4}}, \bibinfo{pages}{6050} (\bibinfo{year}{2014}).

\bibitem{AD1}
\bibinfo{author}{Liu, M.}, \bibinfo{author}{Zhang, J.}, \bibinfo{author}{Adeli,
  E.} \& \bibinfo{author}{Shen, D.}
\newblock \bibinfo{journal}{\bibinfo{title}{Joint classification and regression
  via deep multi-task multi-channel learning for alzheimer's disease
  diagnosis}}.
\newblock {\emph{\JournalTitle{IEEE Transactions on Biomedical Engineering}}}
  \textbf{\bibinfo{volume}{66}}, \bibinfo{pages}{1195--1206}
  (\bibinfo{year}{2018}).

\bibitem{face-temperature}
\bibinfo{author}{Sonkusare, S.} \emph{et~al.}
\newblock \bibinfo{journal}{\bibinfo{title}{Detecting changes in facial
  temperature induced by a sudden auditory stimulus based on deep
  learning-assisted face tracking}}.
\newblock {\emph{\JournalTitle{Scientific reports}}}
  \textbf{\bibinfo{volume}{9}}, \bibinfo{pages}{4729} (\bibinfo{year}{2019}).

\bibitem{cat}
\bibinfo{author}{Finka, L.~R.} \emph{et~al.}
\newblock \bibinfo{journal}{\bibinfo{title}{Geometric morphometrics for the
  study of facial expressions in non-human animals, using the domestic cat as
  an exemplar}}.
\newblock {\emph{\JournalTitle{Scientific reports}}}
  \textbf{\bibinfo{volume}{9}}, \bibinfo{pages}{9883} (\bibinfo{year}{2019}).

\bibitem{HyperFace}
\bibinfo{author}{Ranjan, R.}, \bibinfo{author}{Patel, V.~M.} \&
  \bibinfo{author}{Chellappa, R.}
\newblock \bibinfo{journal}{\bibinfo{title}{Hyperface: A deep multi-task
  learning framework for face detection, landmark localization, pose
  estimation, and gender recognition}}.
\newblock {\emph{\JournalTitle{IEEE Transactions on Pattern Analysis and
  Machine Intelligence}}} \textbf{\bibinfo{volume}{41}},
  \bibinfo{pages}{121--135} (\bibinfo{year}{2019}).

\bibitem{DAN}
\bibinfo{author}{Kowalski, M.}, \bibinfo{author}{Naruniec, J.} \&
  \bibinfo{author}{Trzcinski, T.}
\newblock \bibinfo{title}{Deep alignment network: A convolutional neural
  network for robust face alignment}.
\newblock In \emph{\bibinfo{booktitle}{Proceedings of the IEEE Conference on
  Computer Vision and Pattern Recognition Workshops}}, \bibinfo{pages}{88--97}
  (\bibinfo{year}{2017}).

\bibitem{ultrasound_1}
\bibinfo{author}{Vezzetti, E.}, \bibinfo{author}{Speranza, D.},
  \bibinfo{author}{Marcolin, F.}, \bibinfo{author}{Fracastoro, G.} \&
  \bibinfo{author}{Buscicchio, G.}
\newblock \bibinfo{journal}{\bibinfo{title}{Exploiting 3d ultrasound for fetal
  diagnostic purpose through facial landmarking}}.
\newblock {\emph{\JournalTitle{Image Analysis \& Stereology}}}
  \textbf{\bibinfo{volume}{33}}, \bibinfo{pages}{167--188}
  (\bibinfo{year}{2014}).

\bibitem{ultrasound_2}
\bibinfo{author}{Vezzetti, E.}, \bibinfo{author}{Speranza, D.},
  \bibinfo{author}{Marcolin, F.} \& \bibinfo{author}{Fracastoro, G.}
\newblock \bibinfo{journal}{\bibinfo{title}{Diagnosing cleft lip pathology in
  3d ultrasound: a landmarking-based approach}}.
\newblock {\emph{\JournalTitle{Image Analysis \& Stereology}}}
  \textbf{\bibinfo{volume}{35}}, \bibinfo{pages}{53--65}
  (\bibinfo{year}{2016}).

\bibitem{facial-survey}
\bibinfo{author}{Wu, Y.} \& \bibinfo{author}{Ji, Q.}
\newblock \bibinfo{journal}{\bibinfo{title}{Facial landmark detection: A
  literature survey}}.
\newblock {\emph{\JournalTitle{International Journal of Computer Vision}}}
  \textbf{\bibinfo{volume}{127}}, \bibinfo{pages}{115--142}
  (\bibinfo{year}{2019}).

\bibitem{pose-survey}
\bibinfo{author}{Gong, W.} \emph{et~al.}
\newblock \bibinfo{journal}{\bibinfo{title}{Human pose estimation from
  monocular images: A comprehensive survey}}.
\newblock {\emph{\JournalTitle{Sensors}}} \textbf{\bibinfo{volume}{16}},
  \bibinfo{pages}{1966} (\bibinfo{year}{2016}).

\bibitem{dl}
\bibinfo{author}{Pouyanfar, S.} \emph{et~al.}
\newblock \bibinfo{journal}{\bibinfo{title}{A survey on deep learning:
  Algorithms, techniques, and applications}}.
\newblock {\emph{\JournalTitle{ACM Computing Surveys (CSUR)}}}
  \textbf{\bibinfo{volume}{51}}, \bibinfo{pages}{92} (\bibinfo{year}{2019}).

\bibitem{dl_image}
\bibinfo{author}{Voulodimos, A.}, \bibinfo{author}{Doulamis, N.},
  \bibinfo{author}{Doulamis, A.} \& \bibinfo{author}{Protopapadakis, E.}
\newblock \bibinfo{journal}{\bibinfo{title}{Deep learning for computer vision:
  A brief review}}.
\newblock {\emph{\JournalTitle{Computational intelligence and neuroscience}}}
  \textbf{\bibinfo{volume}{2018}} (\bibinfo{year}{2018}).

\bibitem{dl_cnn}
\bibinfo{author}{Gu, J.} \emph{et~al.}
\newblock \bibinfo{journal}{\bibinfo{title}{Recent advances in convolutional
  neural networks}}.
\newblock {\emph{\JournalTitle{Pattern Recognition}}}
  \textbf{\bibinfo{volume}{77}}, \bibinfo{pages}{354--377}
  (\bibinfo{year}{2018}).

\bibitem{heatmap-origin}
\bibinfo{author}{Pfister, T.}, \bibinfo{author}{Charles, J.} \&
  \bibinfo{author}{Zisserman, A.}
\newblock \bibinfo{title}{Flowing convnets for human pose estimation in
  videos}.
\newblock In \emph{\bibinfo{booktitle}{Proceedings of the IEEE International
  Conference on Computer Vision}}, \bibinfo{pages}{1913--1921}
  (\bibinfo{year}{2015}).

\bibitem{300w}
\bibinfo{author}{Sagonas, C.}, \bibinfo{author}{Tzimiropoulos, G.},
  \bibinfo{author}{Zafeiriou, S.} \& \bibinfo{author}{Pantic, M.}
\newblock \bibinfo{title}{300 faces in-the-wild challenge: The first facial
  landmark localization challenge}.
\newblock In \emph{\bibinfo{booktitle}{Proceedings of the IEEE International
  Conference on Computer Vision Workshops}}, \bibinfo{pages}{397--403}
  (\bibinfo{year}{2013}).

\bibitem{andriluka20142d}
\bibinfo{author}{Andriluka, M.}, \bibinfo{author}{Pishchulin, L.},
  \bibinfo{author}{Gehler, P.} \& \bibinfo{author}{Schiele, B.}
\newblock \bibinfo{title}{2d human pose estimation: New benchmark and state of
  the art analysis}.
\newblock In \emph{\bibinfo{booktitle}{Proceedings of the IEEE Conference on
  computer Vision and Pattern Recognition}}, \bibinfo{pages}{3686--3693}
  (\bibinfo{year}{2014}).

\bibitem{ESSV}
\bibinfo{author}{Eslami, M.}, \bibinfo{author}{Neuschaefer-Rube, C.} \&
  \bibinfo{author}{Serrurier, A.}
\newblock \bibinfo{journal}{\bibinfo{title}{Automatic vocal tract segmentation
  based on conditional generative adversarial neural network}}.
\newblock {\emph{\JournalTitle{Studientexte zur Sprachkommunikation:
  Elektronische Sprachsignalverarbeitung 2019}}} \bibinfo{pages}{263--270}
  (\bibinfo{year}{2019}).

\bibitem{staticMRIData}
\bibinfo{author}{Vald\'es~Vargas, J.~A.}
\newblock \emph{\bibinfo{title}{Adaptation of orofacial clones to the
  morphology and control strategies of target speakers for speech
  articulation}}.
\newblock Ph.D. thesis, \bibinfo{school}{Universit{\'e} de Grenoble}
  (\bibinfo{year}{2013}).

\bibitem{CPP}
\bibinfo{title}{Comités de protection des personnes (cpp)}.
\newblock
  \bibinfo{howpublished}{\url{https://www.iledefrance.ars.sante.fr/comites-de-protection-des-personnes-cpp}}.
\newblock \bibinfo{note}{[Online; accessed 3-Oct-2019]}.

\bibitem{RCN}
\bibinfo{author}{Honari, S.}, \bibinfo{author}{Yosinski, J.},
  \bibinfo{author}{Vincent, P.} \& \bibinfo{author}{Pal, C.}
\newblock \bibinfo{title}{Recombinator networks: Learning coarse-to-fine
  feature aggregation}.
\newblock In \emph{\bibinfo{booktitle}{Computer Vision and Pattern Recognition
  (CVPR), 2016 IEEE Conference on}} (\bibinfo{organization}{IEEE},
  \bibinfo{year}{2016}).

\bibitem{dlib}
\bibinfo{author}{Kazemi, V.} \& \bibinfo{author}{Sullivan, J.}
\newblock \bibinfo{title}{One millisecond face alignment with an ensemble of
  regression trees}.
\newblock In \emph{\bibinfo{booktitle}{Proceedings of the IEEE conference on
  computer vision and pattern recognition}}, \bibinfo{pages}{1867--1874}
  (\bibinfo{year}{2014}).

\bibitem{ResNet}
\bibinfo{author}{He, K.}, \bibinfo{author}{Zhang, X.}, \bibinfo{author}{Ren,
  S.} \& \bibinfo{author}{Sun, J.}
\newblock \bibinfo{title}{Deep residual learning for image recognition}.
\newblock In \emph{\bibinfo{booktitle}{Proceedings of the IEEE conference on
  computer vision and pattern recognition}}, \bibinfo{pages}{770--778}
  (\bibinfo{year}{2016}).

\bibitem{shapenet}
\bibinfo{author}{Kopaczka, M.}, \bibinfo{author}{Schock, J.} \&
  \bibinfo{author}{Merhof, D.}
\newblock \bibinfo{journal}{\bibinfo{title}{Super-realtime facial landmark
  detection and shape fitting by deep regression of shape model parameters}}.
\newblock {\emph{\JournalTitle{arXiv preprint arXiv:1902.03459}}}
  (\bibinfo{year}{2019}).

\bibitem{Stacked-hourglass}
\bibinfo{author}{Yang, J.}, \bibinfo{author}{Liu, Q.} \&
  \bibinfo{author}{Zhang, K.}
\newblock \bibinfo{title}{Stacked hourglass network for robust facial landmark
  localisation}.
\newblock In \emph{\bibinfo{booktitle}{Proceedings of the IEEE Conference on
  Computer Vision and Pattern Recognition Workshops}}, \bibinfo{pages}{79--87}
  (\bibinfo{year}{2017}).

\bibitem{MCAM}
\bibinfo{author}{Chu, X.} \emph{et~al.}
\newblock \bibinfo{title}{Multi-context attention for human pose estimation}.
\newblock In \emph{\bibinfo{booktitle}{Proceedings of the IEEE Conference on
  Computer Vision and Pattern Recognition}}, \bibinfo{pages}{1831--1840}
  (\bibinfo{year}{2017}).

\bibitem{u-net}
\bibinfo{author}{Ronneberger, O.}, \bibinfo{author}{Fischer, P.} \&
  \bibinfo{author}{Brox, T.}
\newblock \bibinfo{title}{U-net: Convolutional networks for biomedical image
  segmentation}.
\newblock In \emph{\bibinfo{booktitle}{International Conference on Medical
  image computing and computer-assisted intervention}},
  \bibinfo{pages}{234--241} (\bibinfo{organization}{Springer},
  \bibinfo{year}{2015}).

\bibitem{pix2pix}
\bibinfo{author}{Isola, P.}, \bibinfo{author}{Zhu, J.-Y.},
  \bibinfo{author}{Zhou, T.} \& \bibinfo{author}{Efros, A.~A.}
\newblock \bibinfo{title}{Image-to-image translation with conditional
  adversarial networks}.
\newblock In \emph{\bibinfo{booktitle}{Proceedings of the IEEE conference on
  computer vision and pattern recognition}}, \bibinfo{pages}{1125--1134}
  (\bibinfo{year}{2017}).

\bibitem{pix2pixHD}
\bibinfo{author}{Wang, T.-C.} \emph{et~al.}
\newblock \bibinfo{title}{High-resolution image synthesis and semantic
  manipulation with conditional gans}.
\newblock In \emph{\bibinfo{booktitle}{Proceedings of the IEEE conference on
  computer vision and pattern recognition}}, \bibinfo{pages}{8798--8807}
  (\bibinfo{year}{2018}).

\bibitem{augmentor}
\bibinfo{author}{Bloice, M.~D.}, \bibinfo{author}{Stocker, C.} \&
  \bibinfo{author}{Holzinger, A.}
\newblock \bibinfo{journal}{\bibinfo{title}{Augmentor: an image augmentation
  library for machine learning}}.
\newblock {\emph{\JournalTitle{arXiv preprint arXiv:1708.04680}}}
  (\bibinfo{year}{2017}).

\bibitem{takemoto2004}
\bibinfo{author}{Takemoto, H.}, \bibinfo{author}{Kitamura, T.},
  \bibinfo{author}{Nishimoto, H.} \& \bibinfo{author}{Honda, K.}
\newblock \bibinfo{journal}{\bibinfo{title}{A method of teeth superimposition
  on {MRI} data for accurate measurement of vocal tract shape and dimensions}}.
\newblock {\emph{\JournalTitle{Acoustical Science and Technology}}}
  \textbf{\bibinfo{volume}{25}}, \bibinfo{pages}{468--474}
  (\bibinfo{year}{2004}).

\bibitem{ananthakrishnan2010}
\bibinfo{author}{Ananthakrishnan, G.}, \bibinfo{author}{Badin, P.},
  \bibinfo{author}{Vald{\'e}s~Vargas, J.~A.} \& \bibinfo{author}{Engwall, O.}
\newblock \bibinfo{title}{Predicting unseen articulations from multi-speaker
  articulatory models}.
\newblock In \emph{\bibinfo{booktitle}{Proceedings of {Interspeech} 2010}}
  (\bibinfo{address}{Makuhari, Japan}, \bibinfo{year}{2010}).

\bibitem{regularization}
\bibinfo{author}{Zheng, Q.}, \bibinfo{author}{Yang, M.}, \bibinfo{author}{Yang,
  J.}, \bibinfo{author}{Zhang, Q.} \& \bibinfo{author}{Zhang, X.}
\newblock \bibinfo{journal}{\bibinfo{title}{Improvement of generalization
  ability of deep cnn via implicit regularization in two-stage training
  process}}.
\newblock {\emph{\JournalTitle{IEEE Access}}} \textbf{\bibinfo{volume}{6}},
  \bibinfo{pages}{15844--15869} (\bibinfo{year}{2018}).

\bibitem{book}
\bibinfo{author}{Goodfellow, I.}, \bibinfo{author}{Bengio, Y.} \&
  \bibinfo{author}{Courville, A.}
\newblock \emph{\bibinfo{title}{Deep learning}} (\bibinfo{publisher}{MIT
  press}, \bibinfo{year}{2016}).

\bibitem{deepforest}
\bibinfo{author}{Zhou, Z.-H.} \& \bibinfo{author}{Feng, J.}
\newblock \bibinfo{journal}{\bibinfo{title}{Deep forest: Towards an alternative
  to deep neural networks}}.
\newblock {\emph{\JournalTitle{arXiv preprint arXiv:1702.08835}}}
  (\bibinfo{year}{2017}).

\bibitem{ensemble}
\bibinfo{author}{Orlando, J.~I.}, \bibinfo{author}{Prokofyeva, E.},
  \bibinfo{author}{del Fresno, M.} \& \bibinfo{author}{Blaschko, M.~B.}
\newblock \bibinfo{journal}{\bibinfo{title}{An ensemble deep learning based
  approach for red lesion detection in fundus images}}.
\newblock {\emph{\JournalTitle{Computer methods and programs in biomedicine}}}
  \textbf{\bibinfo{volume}{153}}, \bibinfo{pages}{115--127}
  (\bibinfo{year}{2018}).

\end{thebibliography}

% #########################
% #########################
% #########################
% #########################
% #########################
% #########################
% #########################
% #########################

\section*{Acknowledgements}

The authors are very grateful to P. Badin for providing the data, L. Lamalle for recording them, and J.-A. Vald\'es Vargas and G. Ananthakrishnan for performing the majority of the initial landmark labelling. This research project is supported by the START-Program of the Faculty of Medicine, RWTH Aachen. The data component of this work has been partially funded by the French ANR (grant ANR-08-EMER-001-02 `ARTIS').

% #########################
% #########################
% #########################
% #########################
% #########################
% #########################
% #########################
% #########################

\section*{Author contributions statement}

M.E. and A.S. conceived the study, processed the data and wrote the manuscript, M.E. designed the networks and conducted the experiment, M.E., C.N.-R. and A.S. analyzed the results and reviewed the manuscript.

% #########################
% #########################
% #########################
% #########################
% #########################
% #########################
% #########################
% #########################
\section*{Additional information}

% To include, in this order: \textbf{Accession codes} (where applicable); \textbf{Competing interests} (mandatory statement). 
% The corresponding author is responsible for submitting a \href{http://www.nature.com/srep/policies/index.html#competing}{competing interests statement} on behalf of all authors of the paper. This statement must be included in the submitted article file.

The authors declare no significant competing financial, professional, or personal interests that might have influenced the performance or presentation of the work described in this manuscript.

\section*{Data Availability}

The source codes of the methods are available at \href{https://github.com/mohaEs}{https://github.com/mohaEs}. The trained models are available from the author A.S. on reasonable request. Authors have not a permission to share the data.
%A preprint version of this article is available at: .

% #########################
% #########################
% #########################
% #########################
% #########################
% #########################
% #########################
% #########################

% #########################
% ######################### Figures

% #########################
% ######################### Tables

\end{document}